\documentclass{aa}
\usepackage{graphicx,txfonts}
\usepackage{natbib}

\newdimen\captwidth
\newdimen\figwidth
\newcommand{\gl}{GJ\,436}
\newcommand{\glb}{GJ\,436b}
\newcommand{\mjup}{M_\mathrm{Jup}}

\newcommand{\rd}{\mathrm{d}}

\newcommand{\ttr}{t_\mathrm{tr}}
\begin{document}
\title{Dynamical evolution of the Gliese 436 planetary system}
\subtitle{Kozai migration as a potential source for Gliese 436b's eccentricity}
\author{H. Beust \inst{1} \and X. Bonfils \inst{1} \and G. Montagnier
\inst{2} \and X. Delfosse\inst{1} \and T. Forveille\inst{1}}
\institute{UJF-Grenoble 1 / CNRS-INSU, Institut de
Plan\'etologie et d'Astrophysique de Grenoble (IPAG) UMR 5274,
Grenoble, F-38041, France \and
European Organization for Astronomical Research in the Southern Hemisphere 
(ESO), Casilla 19001, Santiago 19, Chile}
\date{Received; Accepted}
\offprints{H. Beust}
\mail{Herve.Beust@obs.ujf-grenoble.fr}
\titlerunning{Dynamics of the Gliese 436 system}
\authorrunning{H. Beust et al.}
\abstract
{The close-in planet orbiting \gl\ presents a puzzling
  orbital eccentricity ($e\simeq 0.14$) considering its very short 
orbital period. 
  Given the age of the system, this planet should have been
  tidally circularized a long time ago. Many attempts to explain this
  were proposed in recent years, either
  involving abnormally weak tides, or the perturbing action of a
  distant companion.}{In this paper, we address the latter issue based 
  on Kozai migration. We propose that \glb\ was
  formerly located further away from the star and that it underwent a
  migration induced by a massive, inclined perturber via Kozai
  mechanism. In this context, the perturbations by the companion
  trigger high amplitude variations to \glb\ that cause tides to act
  at periastron. Then the orbit tidally shrinks to reach its
  present day location.}
{We numerically integrate the 3-body system including tides and
  General Relativity correction. We use a modified symplectic
  integrator as well a fully averaged integrator. The former is slower
  but accurate to any order in semi-major axis ratio, while the latter
  is first truncated to some order (4$^\mathrm{th}$) in semi-major
  axis ratio before averaging.}
{We first show that starting from the present-day location of
  \glb\ inevitably leads to damping the Kozai oscillations and to rapidly
  circularizing the planet. Conversely, starting from 5-10 times further
  away allows the onset of Kozai cycles. The tides act in peak
  eccentricity phases and reduce the semi-major axis of the
  planet. The net result is a two fold evolution, characterized by two
  phases: a first one with Kozai cycles and a slowly shrinking
  semi-major axis, and a second one once the planet gets out of the
  Kozai resonance characterized by a more rapid decrease. The
  timescale of this process appears in most cases much longer than the
  standard circularization time of the planet 
  by a factor larger than 50.}
{This model can provide a solution to the eccentricity paradox of
  \glb. Depending on the various orbital configurations (mass and
  location o the perturber, mutual inclination\ldots), it can take
  several Gyrs to \glb\ to achieve a full orbital decrease and
  circularization. According to this scenario, we could be witnessing
  today the second phase of the scenario where the semi-major axis is
  already reduced while the eccentricity is still significant. We then
  explore the parameter space and derive in which conditions this
  model can be realistic given the age of the system. This yields
  constraints on the characteristics of the putative companion.}
\maketitle
\section{Introduction}
%
The M-dwarf GJ 436 has been the subject of growing interest in recent
years. This star is known to host a close-in Neptune-mass planet 
\citep[Gl 436b][]{but04} that was furthermore observed to undergo transit
\citep{gil07}.

The monitoring of the transits of \glb\ helped constraining its
orbital solution. Noticeably, it appears to have a significant
non-zero eccentricity: $e=0.14\pm0.1$ \citep{man07}, furthermore
refined to $e=0.14\pm0.01$ \citep{dem07}. This eccentricity is
abnormally high for a small period planet \citep[$P\simeq 2.64\,$days
][]{bal10}.  With such a small orbital period, tidal forces are
expected to circularize the orbit within much less than the present
age of the system \citep[1--10~Gyr][]{tor07}. Tidal forces seem indeed
to be at work in this system. The detection of the secondary transit
of the planet enabled \citet{dem07} to derive a brightness temperature
of $T=712\pm36\,$K, with flux reradiated across the day-side only.
Conversely, a stellar irradiation / thermal re-radiation balance leads
for to an equilibrium temperature for the planet
$T_\mathrm{eq}=642\,$K, assuming $T_\mathrm{eff}=3350$\,K for \gl\ and
zero albedo for the planet. According to \citet{demi07}, the
temperature difference indicates tidal heading of \glb, but this could
alternatively be due to the $8\,\mu$m sampling the atmosphere in a hot
band pass, if the planetary atmosphere does not radiate like a
black-body.

Many theories were proposed in recent years to
account for the residual eccentricity of \glb. The most straightforward one is
that the tides are sufficiently weak and/or the age of the system is small 
enough to allow a regular tidal circularization not to be achieved \emph{yet}.
This idea was proposed by \citet{mar08} for \glb, after \citet{tri00}
had suggested that this accounts for any close-in exoplanet observed
with significant eccentricity. It  
is in fact related to our poor knowledge of the planet's
tidal dissipation $Q_p$. $Q_p$ is a dimensionless parameter related to
the rate of energy dissipated per orbital period by tidal forced oscillations 
\citep{bar09}. The smaller $Q_p$, the more efficient the tidal dissipation is.
To lowest order, the circularization
time-scale $t_\mathrm{circ}$ of an exoplanet reads
\citep{gol66,jack08}
\begin{equation}
t_\mathrm{circ}=\frac{4}{63}\frac{a^{13/2}}{\sqrt{GM_*^3}}{Q_pm_p}{R_p^5}\qquad,
\label{tcirc}
\end{equation}
where $M_*$ is the stellar mass, $a$ is the planet's orbital
semi-major axis, $m_p$ its mass and $R_p$ its radius. Assuming
$Q_p=10^5$, $R=27,600\,$km, $a=0.0287\,$AU, $m=23.2\,M_\oplus$ and
$M=0.452\,M_\odot$ \citep{mar08}, this formula gives
$t_\mathrm{circ}=4.7\times10^7\,$yr, which is obviously less than the
age of the system. But $Q_p$ is very badly constrained. For Neptune,
it is estimated between $\sim 10^4$ and $3.3\times10^5$
\citep{ban92,zha08}. We can thus consider $10^5$ as a standard likely
value for \glb, but \citet{mar08} argues that $Q_p$ could be as high
as a few $10^6$. In this context, assuming the lower bound (1\,Gyr)
for the age of the system, the circularization could not be achieved
yet. This depends however on the starting eccentricity at time
zero. \citet{wis08} performed analytical calculations of energy
dissipation rates for synchronized bodies at arbitrary eccentricities
and obliquities. His concluded that for a large enough starting
eccentricity ($\ga 0.4$, as will be the case below), the
circularization time-scale should be significantly reduced with
respect to Eq.~(\ref{tcirc}). In this context, the conclusions by
\citet{mar08} may no longer hold. This shows also that conclusions
drawn on asymptotic formulas like Eq.~(\ref{tcirc}) must be treated
with caution. Numerical work is required.

Alternatively, many authors proposed that the eccentricity of \glb\ is
sustained by perturbations by an outer massive planet, despite
a standard $Q_p$ value. Such a long period companion was initially
suspected as \citet{man07}'s radial velocity data revealed a linear drift. 
More data did not confirm the trend, which is now believed to be spurious.
\citet{mon09} gives instead observational uppers limits for the 
putative companion that rule out any companion in the Jupiter mass range 
up to a few AUs.

Perturbers can be either resonant or non-resonant. Looking for a
for a planet locked in mean-motion
resonance with \glb\ appears a promising idea, 
as resonant configuration usually trigger larger
eccentricity modulations. This was suggested by \cite{rib08}, who
fitted in the residuals of the radial velocity data a
$\sim5\,M_\oplus$ planet locked in 2:1 mean motion resonance with
\glb. However, this additional planet was invalidated
furthermore. Dynamical calculations by \citet{mar08} showed that this
planet cannot sustain the eccentricity of \glb\ unless
$Q_p\ga10^6$. Moreover, \citet{alon08} showed that a $\sim
5\,M_\oplus$ planet in 2:1 resonance with \glb\ should trigger transit
time variations that should have been detected yet. Such variations
are unseen today with a high level of confidence
\citep{pont09}. Actually the constraints deduced from radial velocity
monitoring, transit time and geometry monitoring
\citep{bal10,mon09} almost rule out an
additional large planet locked in a low-order mean-motion resonance with
\glb\ ($P<8.5\,$days), and to have a Jovian like planet up to a few
AUs. 

Conversely, \citet{tong09} investigated the effects of 
secular planetary perturbations 
in eccentricity pumping, considering both non-resonant (secular) and
resonant configurations. They found some perturbers configurations
that could account for the present eccentricity of \glb\ while still
being compatible with the observational constraints, but with no
tides. Incorporating the tides leads inevitably to damp all the eccentricity 
modulation and to circularize the orbit. Meanwhile,
\citet{bat09} reanalyzed carefully this issue. They found that as
expected, in most cases the eccentricity of \glb\ quickly falls to
zero thanks to tidal friction, but that this effect can be
considerably delayed if the 2 planet system lies initially in a
specific configuration where the eccentricities of the two planets are
locked at stationary points in the secular dynamics diagram. In
this case, they show that the circularization time can be as high as
$\sim 8\,$Gyr despite a standard $Q_\mathrm{p}$ value.

To date the model by \citet{bat09} is the only one able to explain the
high present day eccentricity of \glb\ with a standard $Q_\mathrm{p}$
value. This model could appear as not generic, as it requires a
specific initial configuration, but \citet{bat11} showed that in the
framework of Hamiltonian planetary dynamics with additional dissipative
forces (which is the case here), these points tend to behave as
attractors. Hence various initial configurations can lead to reach
such a point with a further dynamical evolution like described by
\citet{bat09}. As is shown below, in the model presented here, we
exactly encounter such a configuration where many different routes can
lead to such a stationary point (the transition between Phase 1 and
Phase 2, see Sect. 4).

The purpose of this paper is to present an alternate model based on
Kozai mechanism, assuming again a distant perturber. 
This model can be viewed as complementary to the \citet{bat09} study.
Kozai mechanism
is a major dynamical effect in non-coplanar systems that can trigger
eccentricity modulations up to very high values. We
make a review on this effect, without and with tidal effects taken into 
account and describe our approach in Sects.~2 and 3 respectively. 
In Sect.~4, we
present an application to the case of \gl. We first present
calculations starting from the present day orbit of \glb, with the
result that even Kozai mechanism cannot overcome the damping effect of
tidal friction. Then we present other simulations where \glb\ is
initially put much further away from the star than today. We show that
Kozai mechanism pumps \glb's eccentric regularly up to high values
where tides become active at periastron. The result is a decay of the
orbit that drives it to its present day location but on a time scale
that can be considerably longer than the standard circularization
time. Hence even after several Gyrs, \glb's eccentricity can still be
significant. In Sect.~5, we describe a parametric study of this
scenario and derive in which conditions it is compatible with the
present day situation of \glb. Our conclusions are presented in
Sect.~6.
\section{Kozai mechanism}
\subsection{General features}
Kozai mechanism was first described by \citet{koz62}, initially for
comets on inclined orbit ($\ga 50\degr$) with respect to the ecliptic.
Under the effect of secular planetary perturbations (mainly arising
from Jupiter), their orbit is subject to a periodic evolution that
drives it to lower inclination but very high eccentricity, while the
semi-major axis remains constant. As pointed out by \citet{bai92},
this mechanism is responsible for the origin of most sun-grazing
comets in our Solar System. In the circular restricted 3-body problem,
the Kozai Hamiltonian that describes this secular evolution is
obtained by a double averaging of the original Hamiltonian over the
orbital motions of the planet and of the particle \citep{kin99}. 

\begin{figure}
\includegraphics[angle=-90,width=\columnwidth]{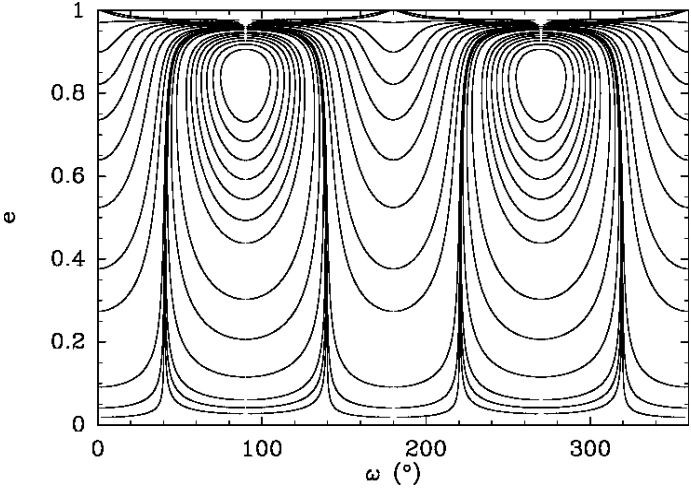}
\caption[]{Contour plots of the Kozai Hamiltonian 
(\ref{f2c}) in a $(\omega,e)$ space for $H=0.05$}
\label{u2c}
\end{figure}
Assuming zero eccentricity for the primaries, the secular 
motion of the particle is characterized, to quadrupolar
approximation (expansion
of the secular Hamiltonian to $2^\mathrm{nd}$ order in powers of the 
semi-major axis ratio) by the following constants of motions
\citep{kin99,kry99,ford00,ggdyn}
\begin{eqnarray}
F & = &(2+3e^2)(3\cos^2i-1)+15e^2\sin^2i\cos2\omega\label{fo2}\\
H & = & (1-e^2)\cos^2 i\qquad,
\end{eqnarray}
where $e$ stands for orbital the eccentricity of the particle, $i$ for
its inclination with respect to the orbital plane of the primaries and 
$\omega$ for the particle's argument of
periastron relative to this plane. The first
constant ($F$) is the reduced Hamiltonian itself and $H$ is the
reduced vertical component of the angular momentum which is a
secular constant thanks to the axial symmetry of the averaged problem. 
Combining both we can eliminate the inclination
\begin{equation}
F=(2+3e^2)\left(\frac{3H}{1-e^2}-1\right)+15e^2\left(1-\frac{H}{1-e^2}\right)
\cos2\omega
\label{f2c}
\end{equation}
Contour plots of this expression in $(\omega,e)$ space for a given $H$
value are called Kozai diagrams. The Kozai mechanism regime is
characterized by small values of $H$ (high initial
inclinations). Figure~\ref{u2c} shows such a diagram corresponding to
$H=0.05$. We note the two islands of libration around $\omega=90\degr$
and $\omega=270\degr$ (or $-90\degr$) 
characteristic for Kozai resonance. In a pure
3-body Newtonian dynamics, the particle just secularly moves along one
of these curves. It is then obvious to see that its evolution is
characterized by strong periodic eccentricity increases. In the same
time $e$ increases, $i$ decreases to keep $H$ constant. If for a given
particle $\omega$ librates around $\pm90\degr$, it is said to be
trapped in Kozai \emph{resonance}. But even if $\omega$ circulates,
its evolution is characterized by the same eccentricity / inclination
modulation described above, the maximum $e$ values being reached when
$\omega=\pm90\degr$.

As quoted above, this regime is valid for small $H$ values. If the
particle has initial zero eccentricity, it can be shown analytically
\citep{kin99} that the minimum inclination required for Kozai
mechanism to start is
$\arccos(\sqrt{3/5})\simeq39.23\degr$. Alternatively, it can operate
at smaller inclination if the initial eccentricity is larger. But a
some point in the secular evolution, it will reach an inclination
above this threshold.

This dynamical mechanism is also active in
hierarchical triple systems \citep{har68,sod82,egg98,kry99,ford00}, the
inner orbit being subject to a similar evolution under secular
perturbations by the outer body. This is particularly relevant for
triple systems because they are often non-coplanar. Similarly, it is
expected to be active on planetary orbits in binary star systems, as
soon as the orbit of the planets are significantly non coplanar with
that of the binary ($i\ga 40\degr$, see above).
This can affect the stability of the planetary
orbits, especially if several planets are present \citep{inn97,malm07}.
It was actually invoked to explain the high eccentricity of some extrasolar
planets in binary systems \citep{hol97, lib09}. 
\subsection{Kozai migration}
This mechanism could also apply to the \gl\ case, as a way to sustain
the eccentricity of \glb. But tides inevitably lead to
a circularization within the standard timescale \citep[see simulations
below and also application to the TyCra system in][]{beu97}.
 
To render Kozai mechanism active despite the presence of tides, one must
lower the strength of the tides. This can be done assuming a
higher $Q_p$ as suggested by \citet{mar08} (see Introduction), but this 
is even not enough. Even with an arbitrary large $Q_p$, there is 
still forced apsidal precession due to the planetary and stellar 
quadrupole moments and to General Relativity (see Sect. 3). With 
the present day semi-major axis of \glb, this is enough to override 
the Kozai mechanism.

Then only way to damp all these effects to allow the onset of Kozai 
cycles is to assume a larger
semi-major axis for the planet during a first evolutionary phase. 
The purpose of the present paper is to
investigate this idea. Of course, the present orbital semi-major axis of
\glb\ is well constrained from radial-velocity monitoring, but our
idea is to suggest that it used to be larger in the past, typically
5--10 times larger than today. In this context, all perturbing 
effects listed are too weak to
prevent the onset of Kozai cycles. Then when the eccentricity increases,
the periastron drops, because as long as tides are not active, 
the semi-major axis remains a secular
invariant. Hence tidal friction starts at periastron. This leads to a
gradual decrease of the orbit (i.e., the semi-major axis) 
which causes it to shrink and to finally
reach its present location. This process is called Kozai migration. It
has been described by \citet{egg01,wu03} and more recently by
\citet{fab07}. It is viewed today as one of the various migration
mechanism that can act in planetary systems. Our aim is to investigate
this scenario in the context of \glb. We show that although it
cannot prevent the final circularization of the planet, under suitable
assumptions it can considerably slow down the circularization process,
thus explaining how the eccentricity of \glb\ could still be
significant a few Gyrs after the formation of the system 
in spite of strong tidal friction.
\section{Tidal forces and $N$-body integration}
\subsection{Basic tidal forces}
Our goal is to test the scenario of Kozai migration to the
\glb\ case. This needs to be done via numerical integration.
 The various tidal forces are
described in \citet{mar02}. We recall briefly this theory here. Let us
consider i) a star with mass $m_s$, radius $R_s$, an apsidal constant
$k_s$ and a dissipation parameter $Q_s$, ii) a planet with mass $m_p$,
radius $R_p$, an apsidal constant $k_p$ and a dissipation parameter
$Q_p$. The tidal forces acting on the planet are
\begin{enumerate}
\item the acceleration due to the quadrupole moment of the planet due to the
  distorsion produced by the presence of the star:
\begin{eqnarray}
\vec{\gamma}_\mathrm{QD,p} & = & \frac{R_p^5\left(1+m_s/m_p\right)k_p}{r^4}
\left\{\left[5\left(\vec{\Omega}_p\cdot \vec{u}\right)^2-\Omega_p^2
-\frac{12Gm_s}{r^3}\right]\vec{u}\right.\nonumber\\
&&\qquad\qquad\qquad\qquad\qquad\left.\rule[-3truemm]{0truemm}{6truemm}
-2\left(\vec{\Omega}_p\cdot \vec{u}\right)
\vec{\Omega}_p\right\}\;,
\end{eqnarray}
where $\vec{\Omega}_p$ is the spin vector of the planet, $r$ is the
distance between the planet and the star, and $\vec{u}=\vec{r}/r$, a
unit vector parallel to the radius vector of the orbit;
\item the acceleration produced by the tidal damping of the planet
\begin{equation}
\vec{\gamma}_\mathrm{TF,p}=-\frac{6nk_pm_s}{Q_pm_p}\,
\left(\frac{R_p}{a}\right)^5\left(\frac{a}{r}\right)^8
\left[2\left(\vec{u}\cdot\dot{\vec{r}}\right)\vec{u}+\dot{\vec{r}}
-r\vec{\Omega}_p\times\vec{u}\right],
\end{equation}
where $a$ is the semi-major axis and $n$ the mean motion of the orbit.
\end{enumerate}
Of course we have symmetric accelerations $\vec{\gamma}_\mathrm{QD,s}$
and $\vec{\gamma}_\mathrm{TF,s}$ for the star. We must add to these
forces the relativistic post-newtonian acceleration
\begin{eqnarray}
\vec{\gamma}_\mathrm{rel} & = &-\frac{G\left(m_p+m_s\right)}{r^2c^2}\left\{\left[
(1+3\eta)\dot{\vec{r}}\cdot\dot{\vec{r}}-2(2+\eta)
\frac{G\left(m_p+m_s\right)}{r}\right.\right.\nonumber\\
&&\qquad\qquad\qquad\qquad\left.\left.\rule[-5truemm]{0truemm}{10truemm}
-\frac{3}{2}\eta\dot{r}^2\right]\vec{u}-2(2-\eta)\dot{r}\,
\dot{\vec{r}}\right\}\;,
\end{eqnarray}
where $\eta=m_pm_s/(m_p+m_s)^2$ and $c$ is the speed of
light. Finally, the equation of motion describing the evolution of the
orbit of the planet is
\begin{equation}
\ddot{\vec{r}}=-\frac{G\left(m_s+m_p\right)}{r^3}\,\vec{r}
+\gamma_\mathrm{rel}+\vec{\gamma}_\mathrm{QD,p}+\vec{\gamma}_\mathrm{TF,p}
+\vec{\gamma}_\mathrm{QD,s}+\vec{\gamma}_\mathrm{TF,s}
\end{equation}
We must also add an equation governing the evolution of the spin
vectors. For the planet we have \citep{mar02}
\begin{equation}
I_p\dot{\vec{\Omega}_p}=-\frac{m_sm_p}{m_s+m_p}\,\vec{r}\times
\left(\vec{\gamma}_\mathrm{QD,p}+\vec{\gamma}_\mathrm{TF,p}\right)\quad,
\label{eqrot}
\end{equation}
and a symmetric equation for $\dot{\vec{\Omega}_s}$.  
Note that we typically have $\gamma_\mathrm{QD,s}\ll\gamma_\mathrm{QD,p}$ and
 $\gamma_\mathrm{TF,s}\ll\gamma_\mathrm{TF,p}$ by a factor $m_p/m_s$, 
showing that the star itself is very little affected. To lowest order for a
synchronized planet, we have
$\gamma_\mathrm{QD,p}/\gamma_\mathrm{TF,p}\simeq11Q_p/6$.
With $Q_p\ga 10^5$. We see
that $\gamma_\mathrm{QD,p}$ dominates the orbital evolution. However,
$\gamma_\mathrm{TF,p}$ (the damping term) is responsible for the
circularization and must be retained. Also, we have to lowest order
\begin{equation}
\frac{\gamma_\mathrm{rel}}{\gamma_\mathrm{QD,p}}\simeq\frac{3}{11k_p}
\frac{\dot{\vec{r}}\cdot\dot{\vec{r}}}{c^2}\frac{r^5}{R_p^5}\frac{m_p}{m_s}
\end{equation}
With the numerical values for the parameters of \glb\ quoted in
\citet{mar08}, this ratio is $\simeq 1.6$. This shows that
relativistic effects are expected to be comparable to tides and must
not be neglected. However, as pointed out by \citet{mar02}, the only
secular effect of the relativistic potential is to produce an apsidal
advance. This also holds for $\vec{\gamma}_\mathrm{QD,p}$ as
 both effects are conservative. They do not affect the circularization 
process due to $\vec{\gamma}_\mathrm{TF,p}$.
\subsection{Inclusion into $N$-body integration and averaging}
As we want to investigate the combined effect of tides and planetary
perturbations, these forces must be included into a $N$-body
integration scheme. Our basic scenario is an inner planet (\glb)
subject to tidal interaction from the star, and perturbed by an outer,
inclined and more massive planet (with mass $m'$ and semi-major axis $a'$)
that is itself not subject to tides.

A major difficulty is that we have to deal with
phenomena  with very different timescales. In the context of \gl, the
period of Kozai cycles is typically $\sim 10^5\,$yr, while the
tidal circularization acts on timescales $\ga10^8\,$yr. Hence an
efficient $N$-body integrator is required.

A standard way to overcome this difficulty is to average all
interaction terms over the orbital motions of both bodies. This can
concern the interaction 3-body Hamiltonian between the planet and the perturber,
as well as the tidal terms. This way only the secular part of the perturbations
is retained.

The averaging of the tides and of the relativistic terms is done only
over the orbital period of the inner planet. To do this, it is
convenient to derive the effect of the tides on the specific angular
momentum vector $\vec{h}=\vec{r}\times\dot{\vec{r}}$ and the
Runge-Lenz vector \citep{mar02,fab07,egg01}
\begin{equation}
\vec{e}=\frac{\dot{\vec{r}}\times\vec{h}}{G\left(m_s+m_p\right)}
-\vec{u}\qquad.
\end{equation}
Each additional acceleration $\vec{\gamma}$ will induce a rate of
change of $\vec{e}$ and $\vec{h}$ given by
\begin{equation}
\frac{\rd\vec{h}}{\rd t}=\vec{r}\times\vec{\gamma}\quad;\quad
\frac{\rd\vec{e}}{\rd t}=\frac{
2\left(\vec{\gamma}\cdot\dot{\vec{r}}\right)\vec{r}
-\left(\vec{r}\cdot\dot{\vec{r}}\right)\vec{\gamma}
-\left(\vec{\gamma}\cdot\dot{\vec{r}}\right)\dot{\vec{r}}}
{G\left(m_s+m_p\right)\quad.}
\label{dhde}
\end{equation}
These equations are then averaged over the orbital period of the
inner planet. Using Hansen coefficients, this can be done in closed
form. Explicit formulas are given by \citet{mar02}.

The averaging of the interaction Hamiltonian with the outer perturber
over the orbital period of both bodies is more complex, as this cannot
be done in closed form. One has to expand the corresponding terms in
ascending powers of the semi-major axis ratio $a/a'$ between the
planet and the perturber (using Legendre polynomials). Assuming
$a/a'\ll 1$, it is truncated to some given order and then averaged
over both orbits. \citet{wu03} and \citet{fab07} truncate the
Hamiltonian to second order (quadrupolar), while \citet{ford00}
recommend third order (octupolar). \citet{mar02} give explicit
formulas up to third order (for coplanar orbits though). As explained
below, the level of accuracy required by our study forced us 
retain terms up to 4$^\mathrm{th}$ order in $a/a'$, i.e. one level
beyond octupolar. We used \textsc{Maple} to derive analytic formulas 
up to this level.

Another way to handle the 3-body interaction is to use symplectic
integration. The main difficulty is to combine it with tidal
forces. 
The theory of symplectic integration  is based on the Hamilton equations of 
motion that apply to any $N$-body problem. Its background is for instance
described in \citet{st92} and \citet{cham99}. The key idea is to split
the Hamiltonian $H$ into pieces, say $H=H_A+H_B$,
each of them being exactly integrable, and then to integrate each part 
separately alternatively. Most common symplectic integrators
\citep{cham99,ld94} are based on the following second order
scheme: At each time-step $\tau$, we must
\begin{enumerate}
\item advance $H_B$ by a half step $\tau/2$;
\item advance $H_A$ for the full time step $\tau$;
\item Re-advance $H_B$ by $\tau/2$.  
\end{enumerate}
In planetary and hierarchical $N$-body problems, $H_A$ corresponds
typically to a collection of independent Keplerian Hamiltonians and
$H_B$ to the disturbing function. We will name
this scheme $B(\tau/2)A(\tau)B(\tau/2)$. Our goal here is to include
tides to it, i.e. dissipative forces. Of course with dissipation the
integration, as well as the real problem, is no longer
symplectic. This does not prevent the integration method from being
efficient, at least for the $N$-body forces. Typically, with the
second order scheme described above, a time-step of $\sim 1/20$ of the
smallest orbital period is enough in standard problems to ensure a
conservation of the energy to $\sim 10^{-6}$ relative level
\citep{beu03}. Including dissipative forces leads to a non
preservation of the energy, but the same time-step can be kept. There
have been actually many successful attempts in recent years to
include dissipative forces into symplectic integrators 
\citep{mal94,keh03,cuk04,mar06}. 

The most straightforward way to proceed is to include the integration
of the various forces given above into the $B(\tau/2)$ sub-steps (the
disturbing part of the integration). However this turns out not to be
efficient in our case. The reason is that the tidal forces are highly
dependant on the distance $r$ between the planet and the star
($\propto r^{-4}$ and even steeper). Therefore, when considering significantly 
eccentric orbits, keeping the standard symplectic time-step leads to an
insufficient mapping of the tidal forces close to periastron. 

To overcome this difficulty, we used averaged formulas only for the tidal 
and relativistic terms. Then we add to our symplectic scheme
two sub-steps $T(\tau/2)$ of integration of these averaged equations for
$\vec{e}$ and $\vec{h}$, in the order 
$B(\tau/2)T(\tau/2)A(\tau)T(\tau/2)B(\tau/2)$:
\begin{enumerate}
\item advance $H_B$ by a half step $\tau/2$;
\item integrate the averaged tidal equations by $\tau/2$;
\item advance $H_A$ for the full time step $\tau$;
\item re-integrate the averaged tidal equations by $\tau/2$;
\item re-advance $H_B$ by $\tau/2$.  
\end{enumerate}
At each $T(\tau/2)$ sub-step, we integrate simultaneously the evolution
of the spin vectors of the bodies via Eq.~(\ref{eqrot}). The advantage
of this approach is that it avoids the use of truncated expansions,
while still being efficient. We modified this way our symplectic
integrator HJS dedicated to hierarchical systems \citep{beu03}. All
computations presented in Sect.~\ref{integs} have been done using this
integrator. Its use is nevertheless more computing time consuming than a
fully averaged integrator. This is why we also built a fully averaged
integrator for our parametric study (Sect.~5).

To check the validity of our integration method, we also made 
some comparisons 
with integrations performed with a conventional Bulirsh \&\ Stoer integrator.
Of course the latter integrations were much more time consuming so that they 
were done over a much less extended time-span. In all cases the agreement 
appeared excellent. An example is shown on Fig.~\ref{1a_short} 
(eccentricity plot) where we have overplotted in red the result of 
the conventional integration. Both curves are almost identical, so that the 
red curve barely appears under the black one.
\section{Application to Gliese 436}
\label{integs}
\subsection{Integration starting from the present state}
\label{presentstate}
\begin{figure*}
\makebox[\textwidth]{
\includegraphics[angle=-90,width=0.49\textwidth]{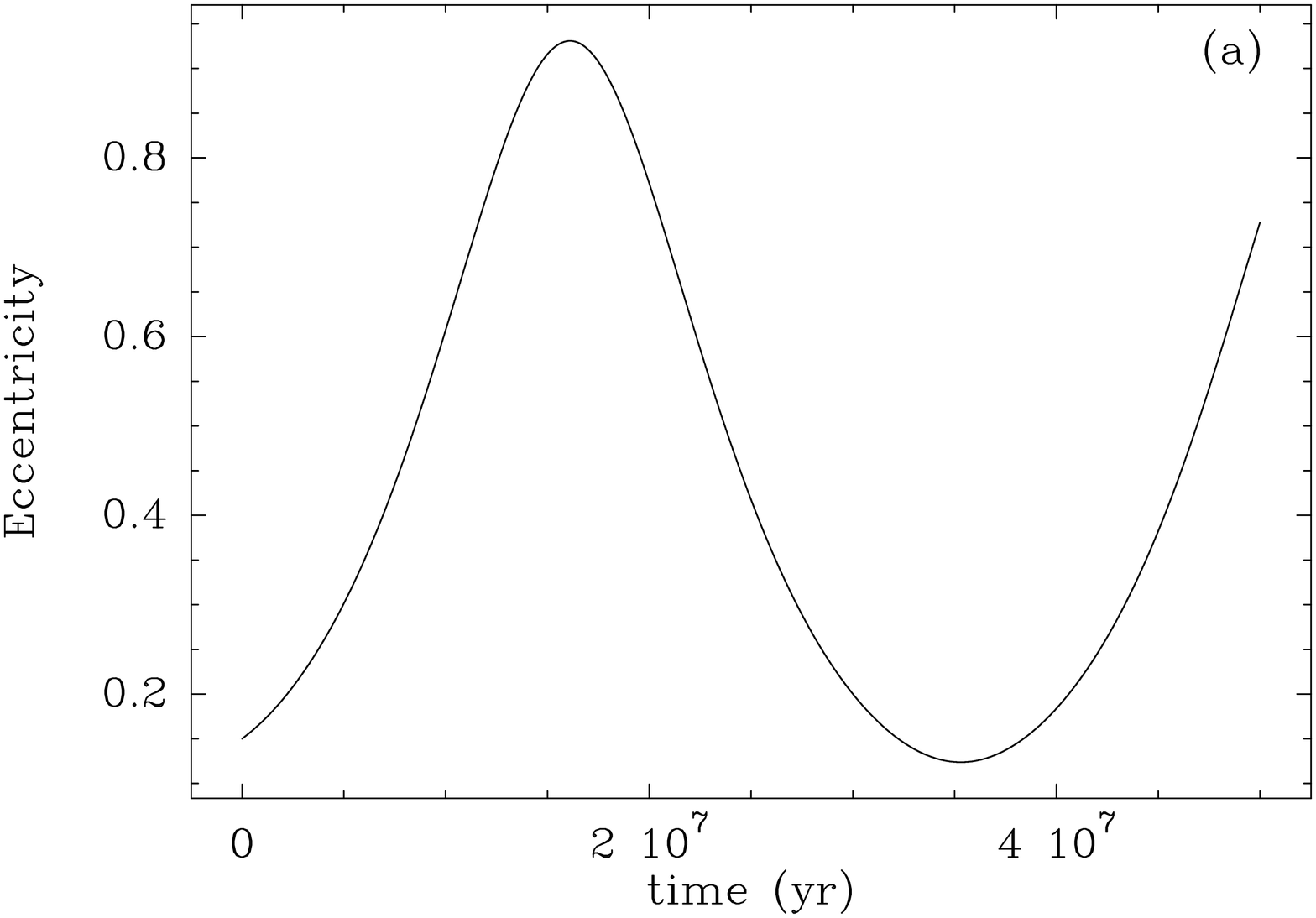} \hfil
\includegraphics[angle=-90,width=0.49\textwidth]{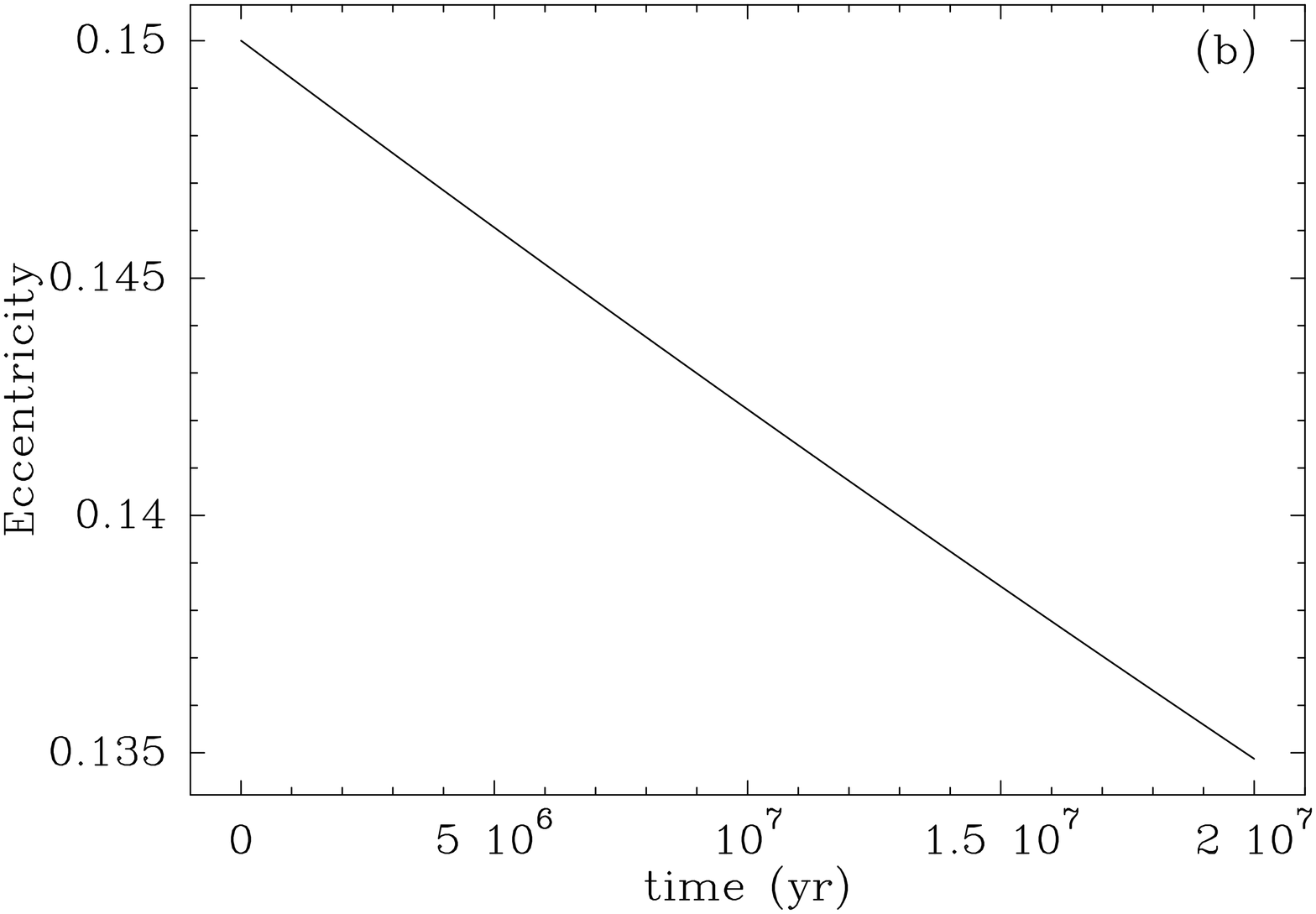}}
\makebox[\textwidth]{
\includegraphics[angle=-90,width=0.49\textwidth]{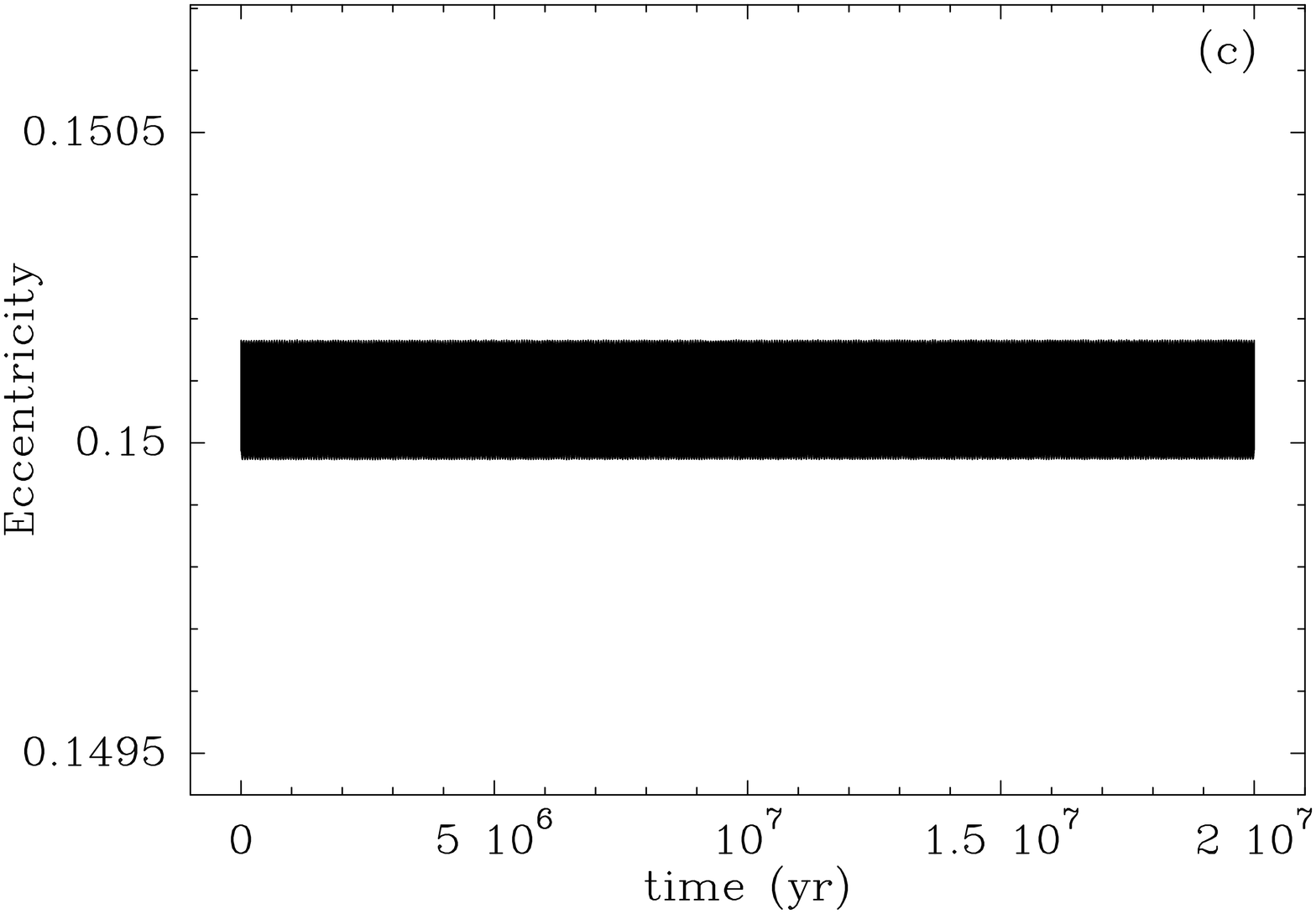} \hfil
\includegraphics[angle=-90,width=0.49\textwidth]{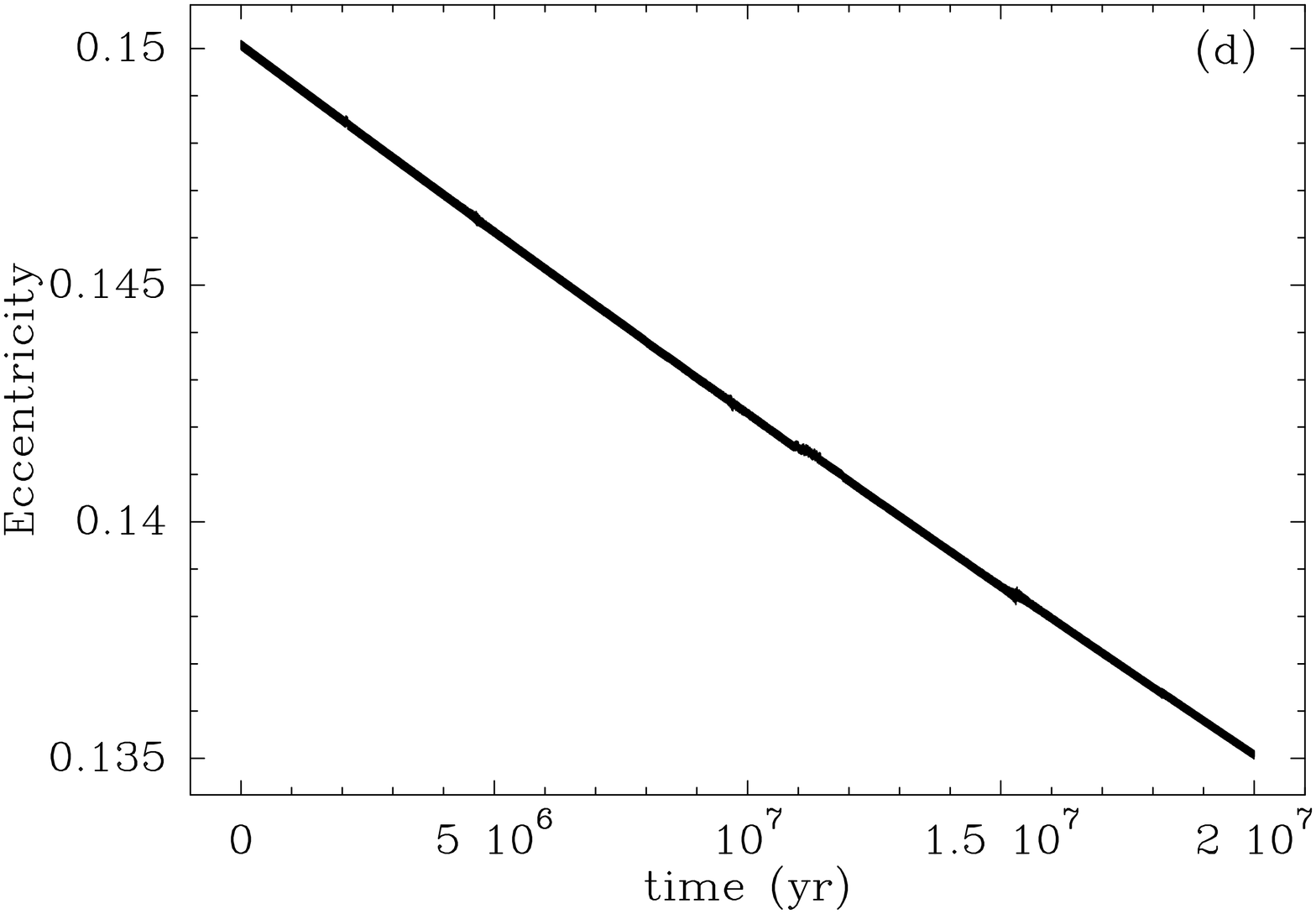}}
\caption[]{Secular evolution of the eccentricity of \glb\ starting
from the present-day orbit and assuming a standard perturber ($m'=0.1 \mjup$,
$T'=10\,$yr) 
in different conditions (see text): (a) perturbing
planet, but no tides nor
relativity; (b) tides and relativity but no perturbing planet; (c)
perturbing planet, relativity but no tides, (d) perturbing planet,
relativity and tides.}  
\label{gl0}
\end{figure*}
The first element we want to test is how planetary perturbations may
affect the present orbit of \glb. As mentioned above, given the
present semi-major axis $a=0.0287\,$AU, tides are expected to dominate
the dynamical evolution. We therefore performed four runs starting
from the present-day orbital configuration with different assumptions: 
\begin{itemize}
\item run (a): with a distant perturber, but not taking into account 
the tides nor the general
relativistic post-newtonian correction (hereafter GR), i.e. 
using the standard HJS code;
\item run (b): taking
into account tides and GR, but with no perturbing planet (run (b) in
Fig.~\ref{gl0});
\item run (c): with the same perturbing planet as in run
(a), but taking now into account the relativistic post-newtonian
correction, but no tides;
\item run (d) : with the same planet taking into account 
both GR and tides.
\end{itemize}
In all cases, the perturbing planet was chosen with
mass $m'=0.1\,\mjup$ and semi-major axis $a'=3.583\,$AU (i.e., with
orbital period $T'=10\,$yr) as a typical convenient set of parameters 
(see Sect.~5), compatible with the constraints of \citet{mon09}
based on radial velocity residuals and
photometry.  In addition we fix the initial tilt angle $i_0$
between the two orbits to $75\degr$ to be able to generate a strong
Kozai resonance. We assume the stellar, planetary and tidal parameters
given by \citet{mar08} and listed in the Introduction.

In Fig.~\ref{gl0} we show the secular evolution of the eccentricity of
\glb\ over $2\times10^7$\,yr in the four cases.  In case (a), we see
as expected a large amplitude modulation characteristic for the Kozai
resonance created by the perturbing planet. In case (b), we note a
gradual circularization with a time-scale $\simeq 2\times 10^8$\,yr.
In case (c) (same as (a) but with GR taken into account), there is no
Kozai resonance anymore. The relativistic precession actually smoothes
the Kozai mechanism. The eccentricity is basically constant with high
frequency, small amplitude oscillations. In run (d) (with tides and
relativity), the high frequency oscillation is superimposed to a tidal
decrease at the same rate as in case (b). The main conclusion we can
derive from these runs is that the perturbing planet could in
principle induce a strong Kozai resonance, but that tides and GR
override this effect. Even GR alone (run (c)) is enough to cancel out
the Kozai resonance. The reason is that the secular precession induced
by GR is larger than that induced by the Kozai resonance. As shown by
\citet{fab07}, GR adds an extra term to the secular precession of the
argument of periastron $\omega$, and subsequently, the maximum
eccentricity of the Kozai cycles is significantly
reduced. This can be understood considering
Eq.~(\ref{f2c}). With an extra precession for $\omega$, the
$\cos2\omega$ term in the averaged Hamiltonian $F$ can be considered
as rapidly varying, so that it secularly averages to zero. Only the
first term can be kept in $F$. This makes the Hamiltonian
independent of all angles, making all related actions constant, and
hence $e$ and $i$ are constant. This is what we get here.

If we consider the dissipative case with tides, the net result in our 
case is that the tidal
circularization process remains almost unchanged (run (d)) with
respect to the case with no perturbing planet (run (b)).

The Kozai resonance appears thus unable to sustain the high present
eccentricity of \glb\ over the age of the system. The only ways to
solve this paradigm are either to increase the strength of the Kozai
resonance either to lower the tides and the relativistic effects. The
former way would imply to increase the mass of the perturbing planet
by at least two orders of magnitude; this would render it incompatible   
with the constraints of \citet{mon09}. The latter way can be achieved
assuming a larger semi-major axis for \glb. Of course the present
semi-major axis is well constrained, but we may assume that it used to
be larger in the past. This is why we come to the idea of Kozai migration.
\subsection{The Kozai migration scenario}
\subsubsection{General description}
\begin{figure*}
\makebox[\textwidth]{
\includegraphics[angle=-90,width=0.49\textwidth]{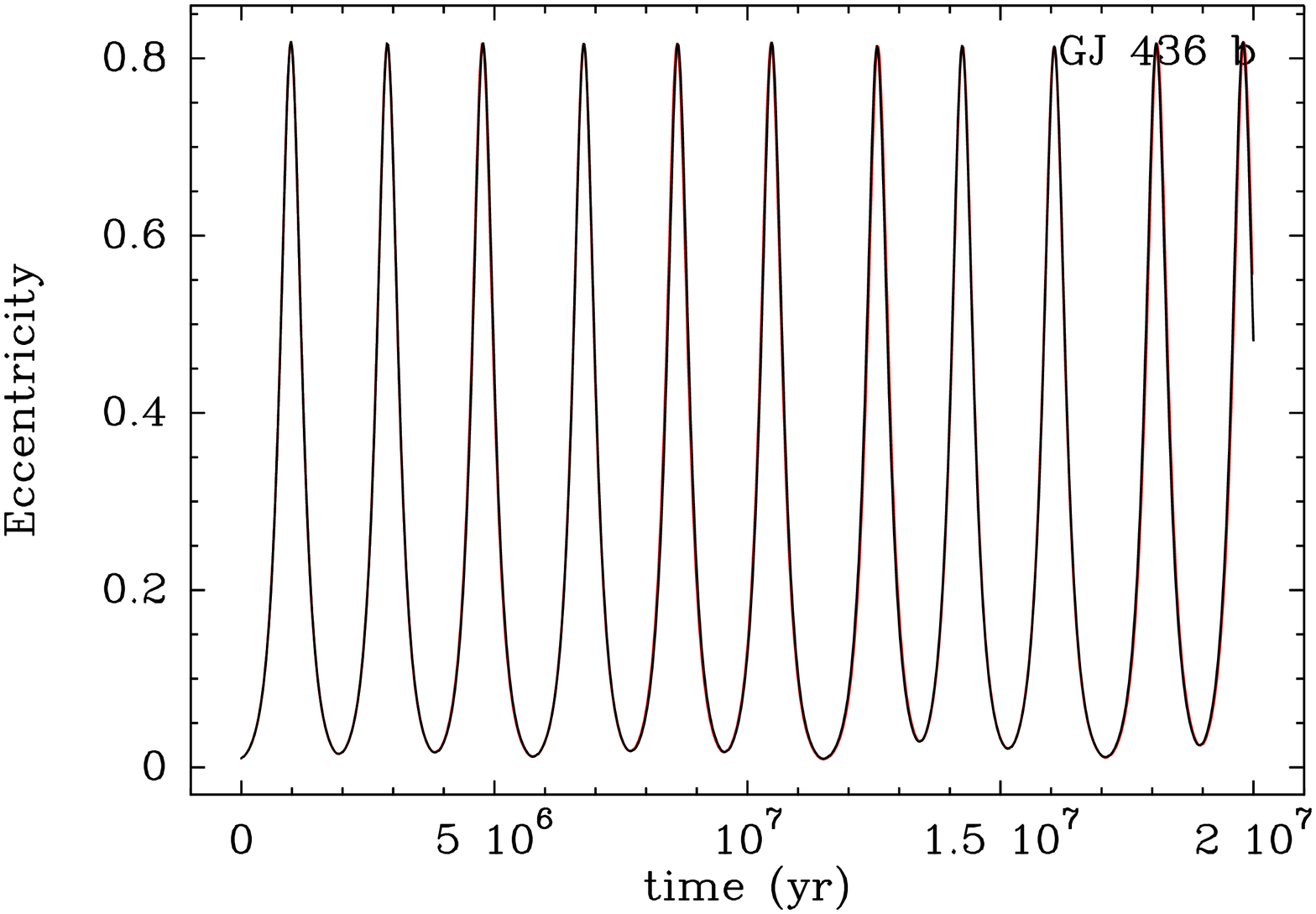} \hfil
\includegraphics[angle=-90,width=0.49\textwidth]{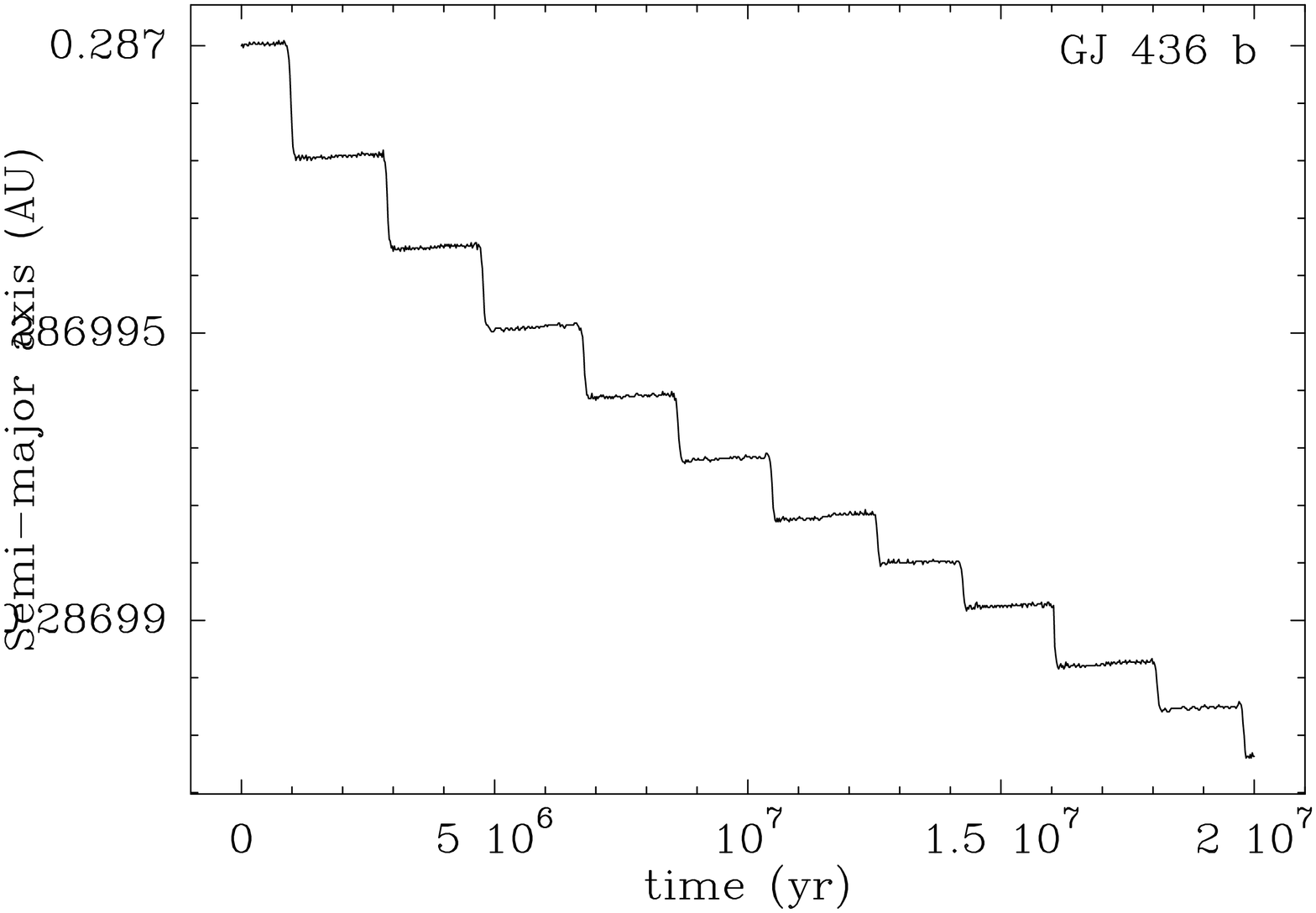}}
\makebox[\textwidth]{
\includegraphics[angle=-90,width=0.49\textwidth]{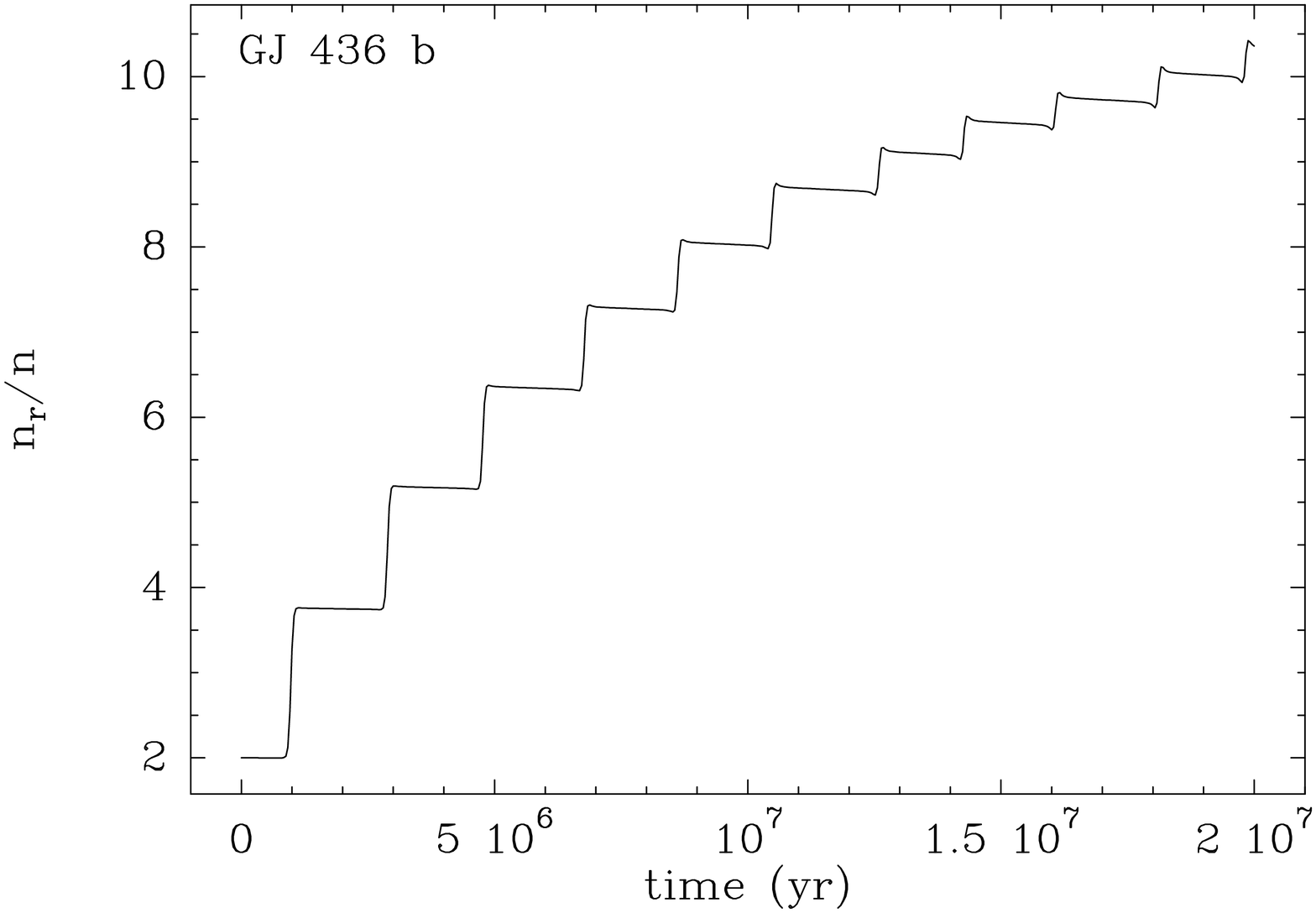} \hfil
\includegraphics[angle=-90,width=0.49\textwidth]{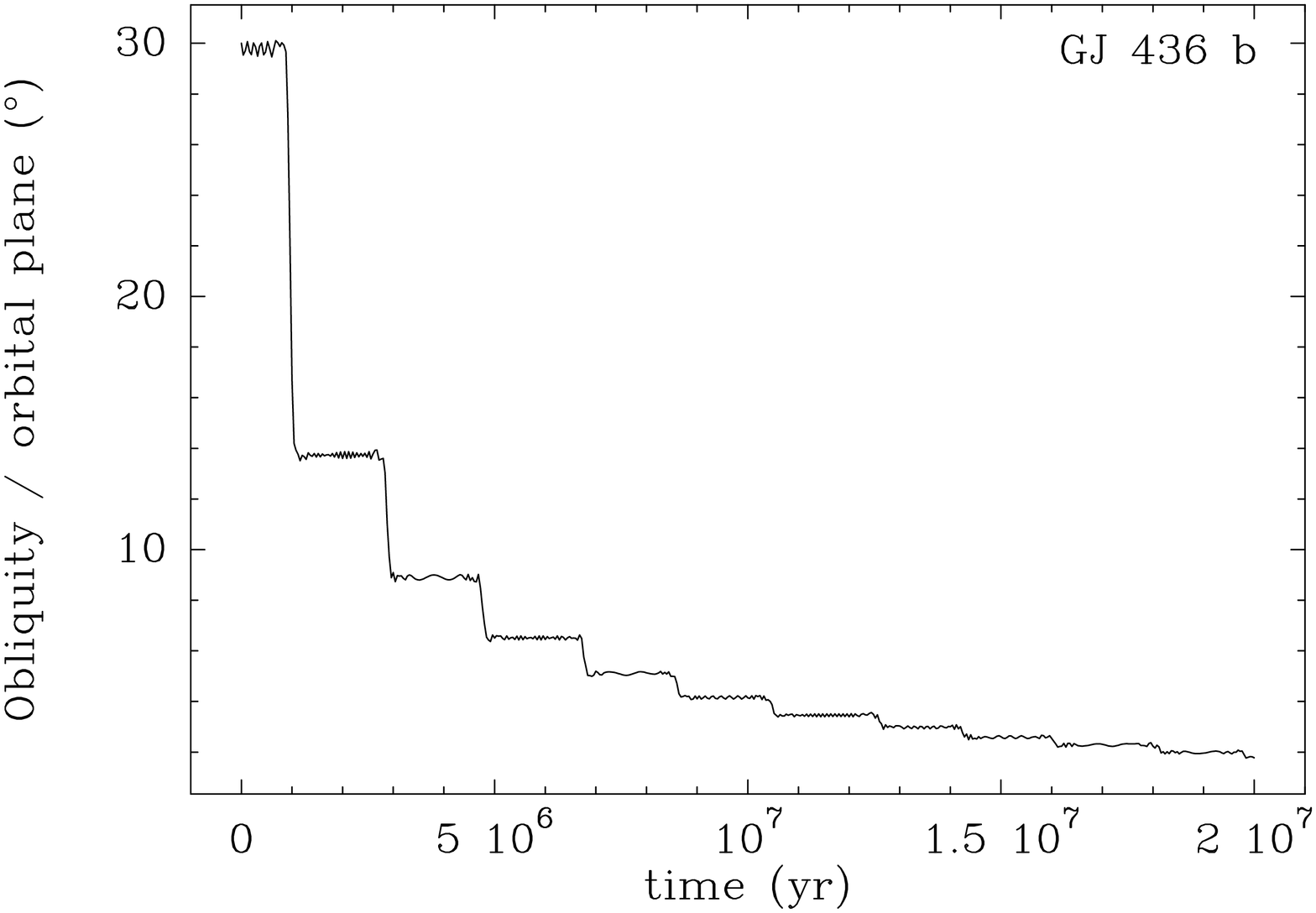}}
\caption[]{Secular evolution over $2\times10^7\;$yr of \glb\ assuming
  an initial value $a=0.287\,$AU and a perturber with $m'=0.1\mjup$
  and $a'=3.583 AU$ ($T'=10\;$yr): Eccentricity (top left), semi-major
  axis (top right), rotation speed with respect to the orbital
  mean-motion ($\Omega/n$, bottom left), and obliquity angle (bottom
  right). On the eccentricity plot we overplot in red the result of
  the same integration, but using a conventional Bulirsh \&\ Stoer
  integrator}
\label{1a_short}
\end{figure*}
\begin{figure*}
\makebox[\textwidth]{
\includegraphics[angle=-90,width=0.49\textwidth]{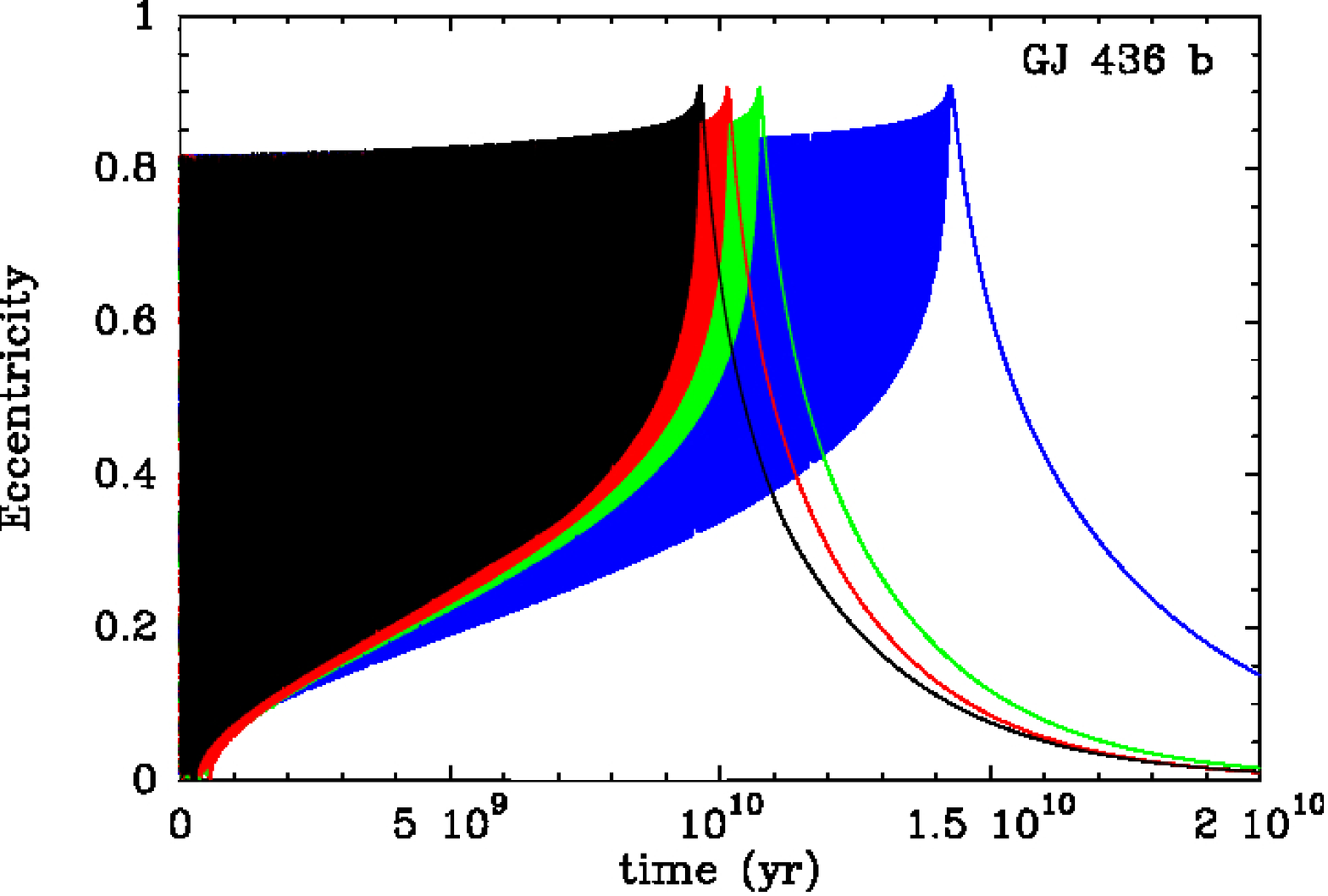} \hfil
\includegraphics[angle=-90,width=0.49\textwidth]{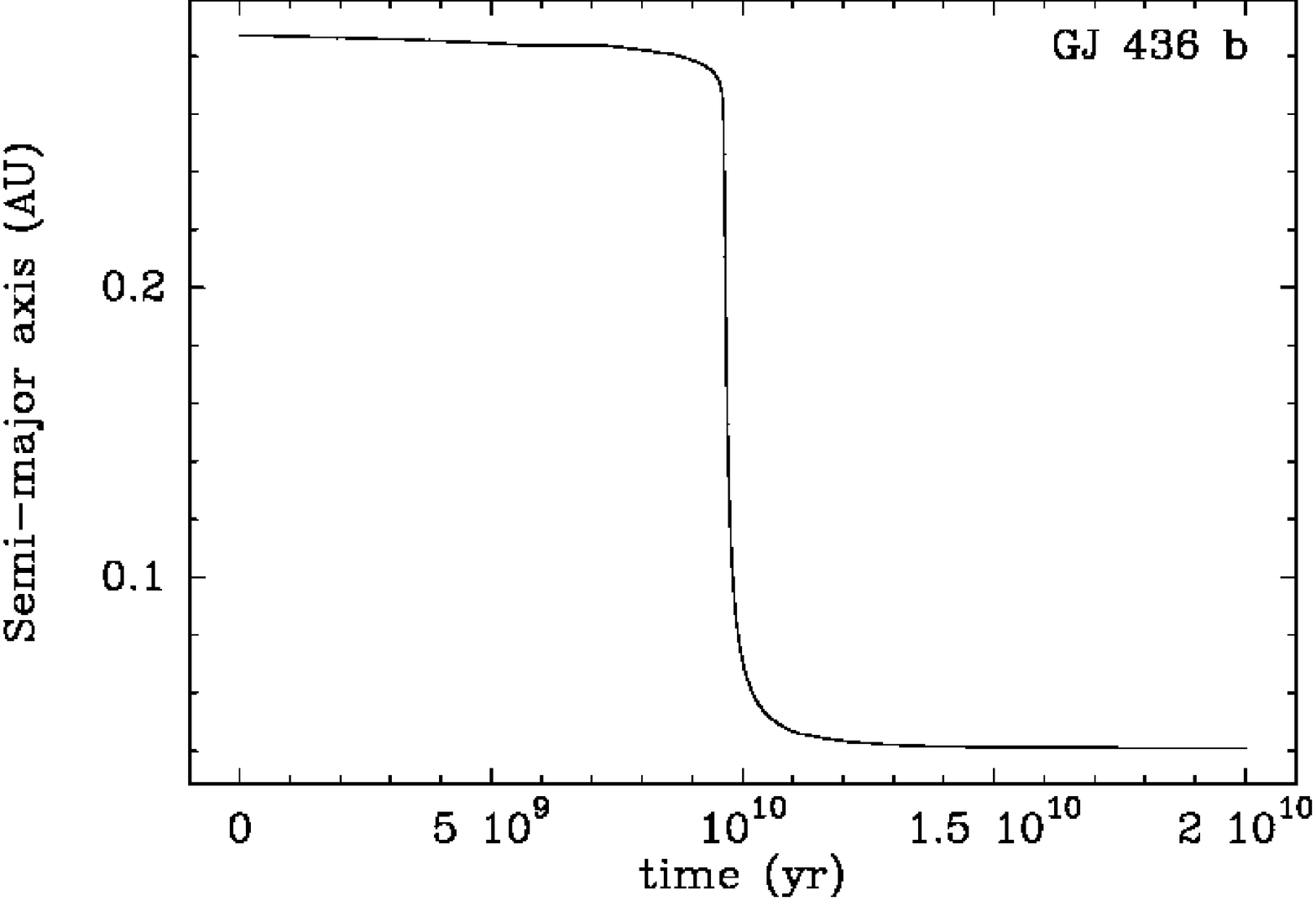}}
\makebox[\textwidth]{
\includegraphics[angle=-90,width=0.49\textwidth]{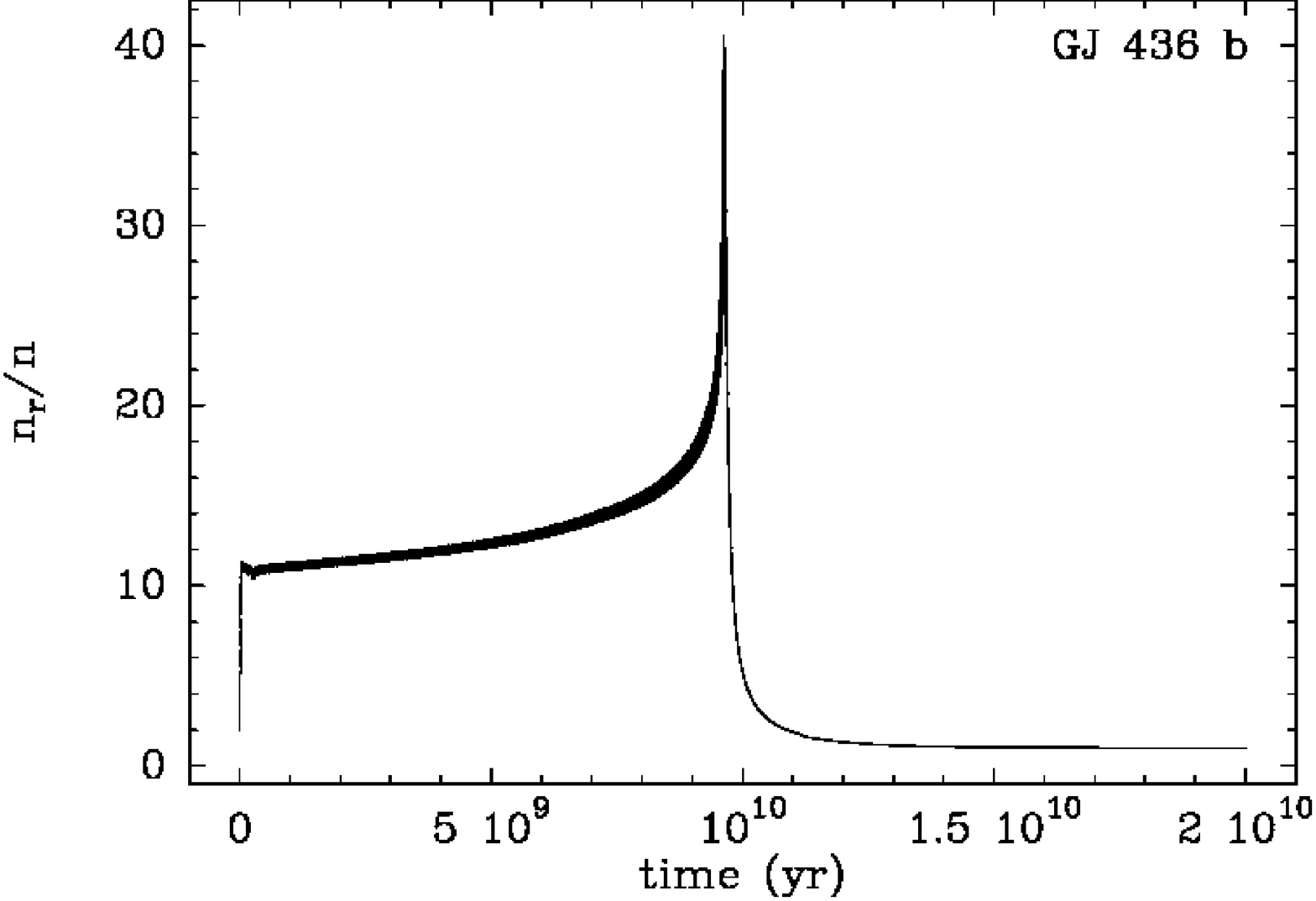} \hfil
\includegraphics[angle=-90,width=0.49\textwidth]{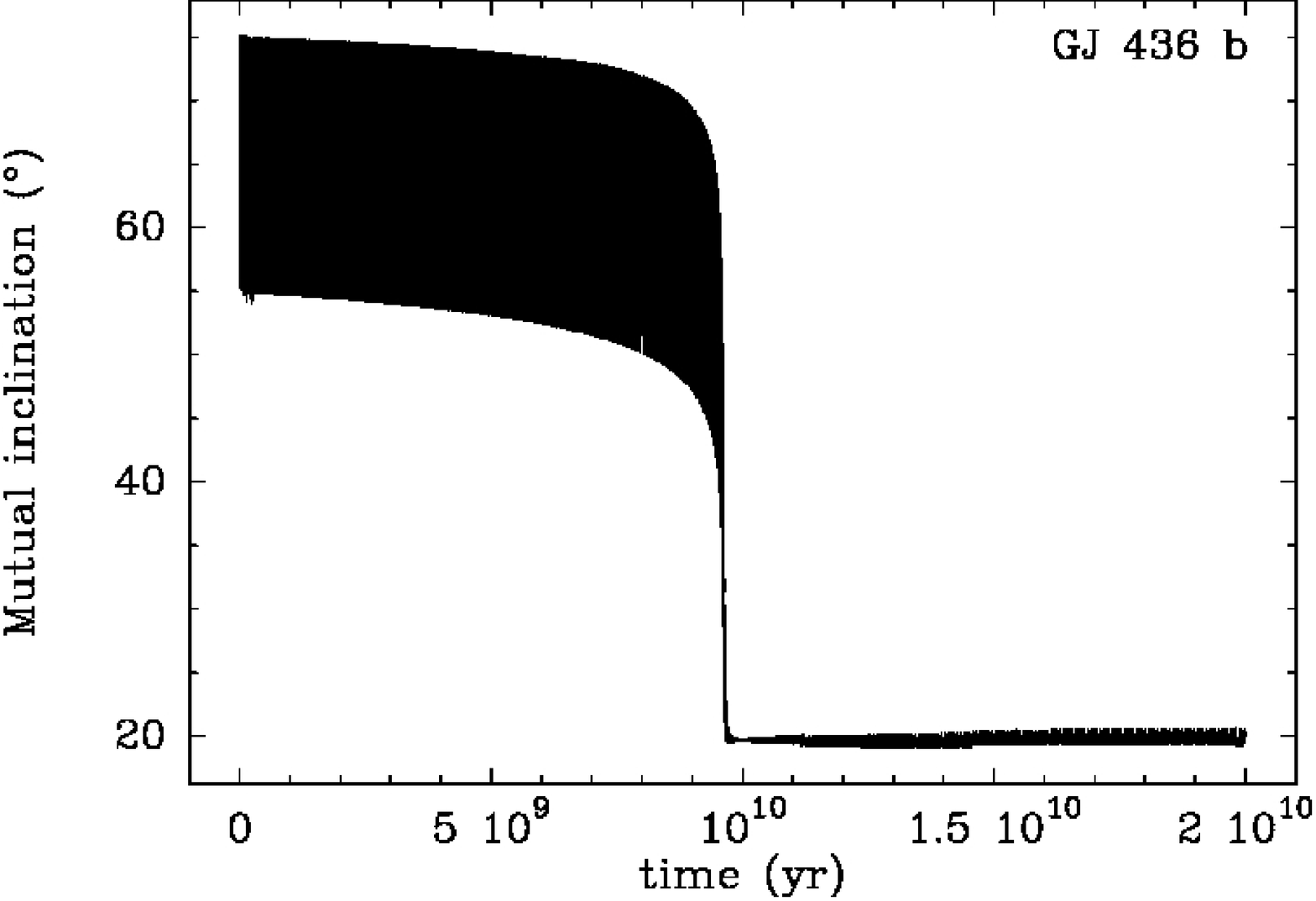}}
\caption[]{Same evolution as in Fig.~\ref{1a_short}, but over
  $2\times10^{10}\;$yr: Eccentricity (top left), semi-major axis (top
  right), rotation speed with respect to the orbital mean-motion
  ($\Omega/n$, bottom left), mutual inclination between the two orbits
  (bottom right). On all plots, the black curve corresponds to our 
  symplectic integration. The color curves on the eccentricity plot are 
calculated with a fully averaged integrator, with interaction Hamiltonian 
truncated up to various orders: 2$^\mathrm{nd}$ order (blue), 
3$^\mathrm{rd}$ order (green), and 4$^\mathrm{th}$ order (red).}
\label{1a_long}
\end{figure*}
We describe now into details a typical run where we assume an
initial semi-major axis for
\glb\ significantly larger than the present value, and where both GR
and tides are taken into account, with unchanged tidal parameters with
respect to Sect.~\ref{presentstate}. We assume for the perturber 
the same parameters as in
Sect.~\ref{presentstate} ($a'=3.583\,$AU, $m'=0.1\,\mjup$, $i_0=75\degr$),
but we take
as initial semi-major axis for \glb\ $a_0=0.287\,$AU (i.e, 10 times
its present-day value). 

Fig.~\ref{1a_short} shows the secular evolution of \glb\ in this context
over $2\times10^7\,$yr. We see that the
eccentricity is subject to a large amplitude modulation characteristic
for Kozai resonance. With $a_0=0.287\,$AU, apsidal precession 
arising from GR and tides is no longer
able to override the Kozai resonance. But at peak eccentricity
($e\simeq0.8$) in the Kozai cycles, the periastron drops to
0.05--0.06\,AU, i.e., only twice the present-day semi-major
axis. As a result, tides are at work in peak eccentricity
phases. This can be seen in Fig.~\ref{1a_short}. We note a rapid
reduction of the obliquity of \glb\ relative to its orbital plane
(initially fixed to $30\degr$), so that after a few $10^7\,$yr, the
spin axis of \glb\ remains perpendicular to its orbital plane. We also
note a spinning up of the planet's rotation. On Fig.~\ref{1a_short}
(bottom-left), we plot as a function of time the ratio of the rotation
velocity \glb\ to its orbital mean-motion ($\Omega/n$), so that 1
means synchronism. This ratio was initially fixed to 2, and we note a
gradual increase up to 10 at $t=2\times10^7\,$yr. This reveals a
spinning up of the planet, as the orbital mean-motion (i.e. the
semi-major axis) is not subject to so drastic changes in the same
time. Finally, we note a very small decrease of the semi-major axis
over the time-span considered. These effects are clearly due to the
tides at work in high eccentricity phases. This shows up obviously in
the staircase-like shape of the temporal evolution of the parameters
displayed. Actually they are subject to some evolution only in the
peak eccentricity phases, and remain unchanged inbetween. The spinning
up of the planet is also a consequence of the tides in high
eccentricity phases. The rotation of \glb\ tries indeed to tidally
synchronize with the orbital angular velocity \emph{at periastron},
which is very different from the mean-motion due to the high
eccentricity. With a peak eccentricity of $\sim0.8$, synchronism with
angular velocity at periastron means actually $\Omega/n\simeq 16$. 
Finally, the semi-major axis decrease reveals a loss of orbital energy
due to tidal friction.

We come now to describing the same simulation, but over a much longer
time span, i.e. 20\,Gyrs (Fig.~\ref{1a_long}). This timescale is 
intentionally taken larger than the age of the system and even 
of the Universe to show the 
possible future evolution. We display as in
Fig.~\ref{1a_short} the eccentricity, the semi-major axis, the
$\Omega/n$ ratio, and the mutual inclination between the two orbits.
On the eccentricity plot, the color curves are 
calculated using various
averaged integrators (see below). The one corresponding to our 
symplectic integration in the black one.
We clearly see a two-fold evolution, first before 10\,Gyrs and then
after. In the first phase, Kozai resonance is still active as can be
seen from the high amplitude eccentricity oscillations. In the same
time, the semi-major axis is subject to a slight decrease. The
eccentricity oscillations do not remain identical though. Actually the
peak eccentricity slightly increases but a more striking fact is that
the bottom eccentricity of the Kozai cycles gradually increases to
finally reach the peak eccentricity after $\sim 10\,$Gyr. At this
point (beginning of the second phase) the eccentricity oscillations
stop, which means that the \glb\ gets out of the Kozai resonance. The
semi-major axis drops then much more quickly and the eccentricity
decreases to zero within a few extra Gyrs. During almost all the
evolution (apart from the beginning), the rotation of \glb\ remains
synchronized with the angular velocity at periastron in high
eccentricity phases. The increase of $\Omega/n$ during the first phase
is related to the gradual increase of the peak eccentricity. In the
second phase $\Omega/n$ gradually decreases towards 1. At the end of
the simulation, \glb\ is tidally synchronized and circularized, but
the final semi-major axis is several times smaller than the initial
one. The Kozai migration is ended. 

This two-fold behaviour in Kozai migration process was already noted
by \citet{wu03} and \citet{fab07}. Actually our results of
Fig.~\ref{1a_long} are very similar to the corresponding ones of
\citet{fab07}.
\subsubsection{The first phase}
During the first phase, the planet still undergoes Kozai oscillations, but the 
bottom eccentricity of the cycles gradually increases to finally reach the 
peak one. This can be explained by a shift in Kozai diagram.

In a pure 3-body Newtonian dynamics, the planet
just secularly moves along one of the level curves in a Kozai diagram 
similar to that of Fig.~\ref{u2c}. In the
$\omega$-circulating case, the curves are explored from left to right
with $\rd\omega/\rd t>0$, and in the $\omega$-librating case (the genuine
``resonant'' orbits), the curves are explored clockwise around the
stable point. Now, both GR and tides add an extra, positive component
to $\rd\omega/\rd t$ \citep{fab07} at peak eccentricity in the
cycles. This results in a slight shift towards right in the Kozai
diagram each time the planet passes through a peak eccentricity phase.
The planet jumps to another nearby curve located to the right. It is
then obvious to see from Fig.~\ref{u2c} that this change induces a
decrease of the \emph{bottom} eccentricity in the $\omega$-librating
case and an increase in the $\omega$-circulating case. Now it is also
obvious that thanks to these shifts in diagram at each eccentricity
peak, a planet initially in the $\omega$-librating case will be pushed
to move to $\omega$-circulating after a while. Consequently the bottom
eccentricity of the planet is expected to eventually first decrease
during the $\omega$-librating phase and then to further increase in
the subsequent $\omega$-circulating regime. This is exactly what is
reported by \citet{fab07} in their Fig.~1. After a short decrease, the
bottom eccentricity of the cycles increases gradually to 1. In our
Fig.~\ref{1a_long}, we see that the bottom eccentricity (black curve)
 of the cycles
does not immediately start to increase, but we do not see an initial decrease
like in \citet{fab07}. In fact, the more realistic integration is
actually ours. The reason is that the integration of \citet{fab07} was
made on the basis of a quadrupolar expansion of the Hamiltonian, as well as
the Kozai diagram in Fig.~\ref{u2c}. It is thus not surprising to see
them in perfect agreement. But our integration, made using a
symplectic integrator, is not truncated and is accurate to any order in
$a/a'$. \citet{ford00} showed that retaining higher order terms, in
particular up to the octupolar level (3$^\mathrm{rd}$ order in
expansion in $a/a'$) considerably improves the accuracy of the
integration, especially in the vicinity of the separatrix between the
two regimes, which is the case discussed here. 

In any case, when the planet enters the $\omega$-circulating regime,
the bottom eccentricity of the cycles is forced to increase.
\subsubsection{The second phase and the link with the \citet{bat09} model}
When the
bottom eccentricity reaches the peak one, then the planet moves along
a more or less straight line in the Kozai diagram. Tides and GR are
now active permanently as the eccentricity is permanently high.
This causes the precession of $\omega$ to
increase. This forces the planet to start a more sharp tidal 
decrease that makes it get out 
of the Kozai resonance. In this second phase, the planet gets gradually 
tidally circularized.

An important outcome of this scenario is that it can
provide a solution to the eccentricity paradox of \glb.
We note in
Fig.~\ref{1a_long} that the first phase of the secular evolution lasts
several Gyrs. Hence \glb\ may have first spent a long time in the
first phase before starting a sharp tidal decrease. Moreover, we note
that in the second phase, even if the semi-major axis shrinks quickly,
it still takes a few more Gyrs to the eccentricity to decrease downto
zero. This could appear surprising, as we are now away from Kozai
resonance. In Fig.~\ref{1a_long}, the situation of the planet at $\sim
13\,$Gyrs is similar to the one we studied in Fig.~\ref{gl0}. In the
latter case, the eccentricity of \glb\ decreases downto zero in $\sim
2\times10^8\,$yr, but in Fig.~\ref{1a_long} it still takes a much longer 
time from 13\,Gyrs to reach zero. How could we explain this discrepancy ?

The difference comes from the dynamical configuration of the 3-body
system. It is in fact closely related to the quasi-stationary solution
described by \citet{bat09}. \citet{bat09} studied the coplanar 3-body
system. They showed that there exist two apsidal fixed points in this
system, characterized by $\rd e/\rd t=0$, and $\varpi-\varpi'=0$
or $\varpi-\varpi'=\pi$,
where $\varpi$ and $\varpi'$ are the longitude of periastron of
\glb\ and the perturber respectively. There is always one stable point
among these two configurations around which librations can occur. In
that case, $\varpi-\varpi'$ just librates around its equilibrium
value, and $\rd e/\rd t=0$ periodically vanishes. As a result, the 
decrease of the eccentricity is much slower. \citet{bat09} showed
that adding tides to this peculiar system considerably increases the
tidal circularization time, because the angular momentum transfer 
associated with tidal dissipation is now shared among the two interacting 
orbits.  This is
of course a peculiar situation, but as mentioned above, these points tend to 
behave as attractors.

The situation we describe here differs from that of \citet{bat09}
because we do not have a coplanar configuration. But the starting
point of the second phase is also characterized by a stationary point
with $\rd e/\rd t=0$. At the end of the first phase (Kozai cycles),
when the bottom eccentricity of the cycles equals the peak one,
\glb\ moves along a straight line in the Kozai diagram. Hence we have
$\rd e/\rd t=0$. Contrary to \citet{bat09}, here $\varpi-\varpi'$ 
circulates during the second phase. The reason is that in the coplanar
problem, there is no quadruplar component to $\rd e/\rd t$. The
expression of $\rd e/\rd t$ given by \citet{bat09} arises indeed from
the octupolar expansion. But at high inclination, the quadrupolar
component to $\rd e/\rd t$ dominates, so that the constraints are
different. But the common feature is that the starting point of both
evolutions is characterized by a stationary point in eccentricity in
the pure secular 3-body evolution. The result is a huge increase of
the tidal circularization time by a factor 50 at least,
like in the evolution described by
\citet{bat09}. Besides, the apsidal alignment criterion could be used 
as a diagnostic. If an apsidally aligned 
companion is discovered in the future in 
the radial velocity residuals of the star,
it would be a strong indication in favor of a coplanar configuration like 
described by \citet{bat09}; conversely if this is not the case, the 
companion could be considered as highly inclined with good probability.
\section{Exploration of the parameter space}
\subsection{Description of the results}
We claim that the Kozai migration process we describe could fit 
the present day configuration of the \gl\ system. \glb\ could fairly 
well be today
in the middle of the second phase, with an already reduced semi-major
axis and a slowly decreasing eccentricity. In Fig.~\ref{1a_long}, the
situation of the planet at $\sim 13\,$Gyrs can be compared to the
present day orbital configuration of \glb. 

Of course 13\,Gyrs is too long compared to the age of the
\gl\ system. But the timescale of the two-fold evolution can vary
considerably depending on the initial parameters of the model, so that
adjusting them to fit to any given age less than the age of the Universe
is possible.

We come now to exploring the parameter space of this scenario to try
to derive in which case it can be compatible with the observation.  We
tested various orbital configurations, exploring the parameter
space. The important parameters are
\begin{itemize}
\item the mass $m'$ and semi-major axis $a'$ (or equivalently the 
orbital period $T'$) of the perturber;
\item the initial semi-major axis $a_0$ of \glb;
\item the initial mutual inclination $i_0$ between the two orbits.
\end{itemize}
Note that instead of $i_0$, the actual critical parameter 
is the reduced vertical 
component of the specific angular momentum 
$h=\sqrt{1-e^2}\,\cos i$, 
which is a secular constant of motion (See Sect.~2.1). But in all our 
simulations we took an initial eccentricity $e_0=0.01$ for \glb, 
so that $h$ is unambiguously related to $i_0$.

Other parameters such as the eccentricity of the perturber, the
initial rotation speed and rotation axis orientation of \glb\ turned
out to have only a minor influence on the global behaviour of the
system, so that exploring the space of the major parameters is enough
for our purpose. In particular, in all the following, the initial
eccentricity $e'$ of the perturber was set to $0.03$ as a typical
standard value.
 
The key outcome to monitor is the transition time between the two
phases in the scenario (hereafter $\ttr$), which occurs close
to 10\,Gyrs in Fig.~\ref{1a_long}. This timescale appeared to be
extremely sensitive to the parameters, ranging from less than
$10^5\,$yr in some extreme cases to hundreds of Gyrs in other
configurations. In all cases, most of the orbital decrease and
circularization that occurs after $\ttr$ is achieved with
$\sim1.5\ttr$.

In our exploration, we consider as likely any $\ttr$ value between
1\,Gyr and 10\,Gyr. Shorter values are not compatible with a not yet
circularized \glb\ today, and with longer values, the star is not old
enough to have entered the second phase yet. Our goal was to derive
which sets of parameters yield a convenient $t_r$. To do this, we
first used our symplectic code in a few cases, and then our averaged
integrator to save computing time. As mentioned above, the averaging
process of the Hamiltonian due to the perturber cannot be done in
closed form.  One has to first expand it in ascending powers of
$a/a'$, to truncate it to some given order and then average over both
orbits. \citet{fab07} truncate to second order (quadrupolar), while
\citet{ford00} recommends third order (octupolar).  We checked that
being in correct agreement with the symplectic integration (in
particular in the monitoring of $\ttr$ which is a very sensitive
parameter) requires to retain terms up to 4$^\mathrm{th}$ order in
$a/a'$, i.e. one level beyond octupolar. The use of a truncated
integrator does not change the global two-fold behaviour reported
above, but it affects its timescale. This is illustrated in 
Fig.~\ref{1a_long}a (eccentricity plot) where we have superimposed to our 
symplectic integration (black curve) the same eccentricity evolution, but 
calculated with averaged integrators truncated at various orders: 
second (blue), third (green) and  4$^\mathrm{th}$ (red). 
The use of the quadrupolar truncation generally leads to an error on 
$\ttr$ of $\sim 50\%$. Most of the time it overestimates it.
This can be understood easily: In a pure Newtonian 
dynamics (no tides, no relativity, such as in Fig.~\ref{gl0}a), all Kozai 
cycles are strictly identical in the quadrupolar approximation. 
The motion is strictly periodic, as can be seen from Fig.~\ref{u2c}. This 
is due to the absence of the perturber's eccentricity and argument of 
periastron in the interaction Hamiltonian (Eq.~(\ref{fo2})). 
This is no longer true
at higher order, as shown by \citet{kry99,ford00}. Due to secular
changes of the perturber's orbit, the Kozai cycles are no longer 
identical (see Fig.~\ref{1a_short}a for instance). The total number 
of Kozai cycles achieved within a given (long) time-span can thus vary 
depending on the theory assumed. As the efficiency of the Kozai migration 
process depends on the integrated number of peak eccentricity phases reached, 
this explains the variability of $\ttr$. We see from Fig.~\ref{1a_long}a 
that the use of a third order (octupolar) theory (green curve) considerably 
enhances the agreement with the symplectic integration. A $\sim 5$\%\ agreement
is reached if we go to 4$^\mathrm{th}$ order (red curve). Or course better 
agreement could be reached if we used even higher order formulas. But in 
that case, the averaged formulas are so complex that the gain in computing 
time with respect to the symplectic integration is small. As a 5\%\ 
accuracy on $\ttr$ is enough for our analysis, we decided to 
use 4$^\mathrm{th}$ order theory.

\begin{figure*}
\makebox[\textwidth]{
\includegraphics[angle=-90,width=0.49\textwidth]{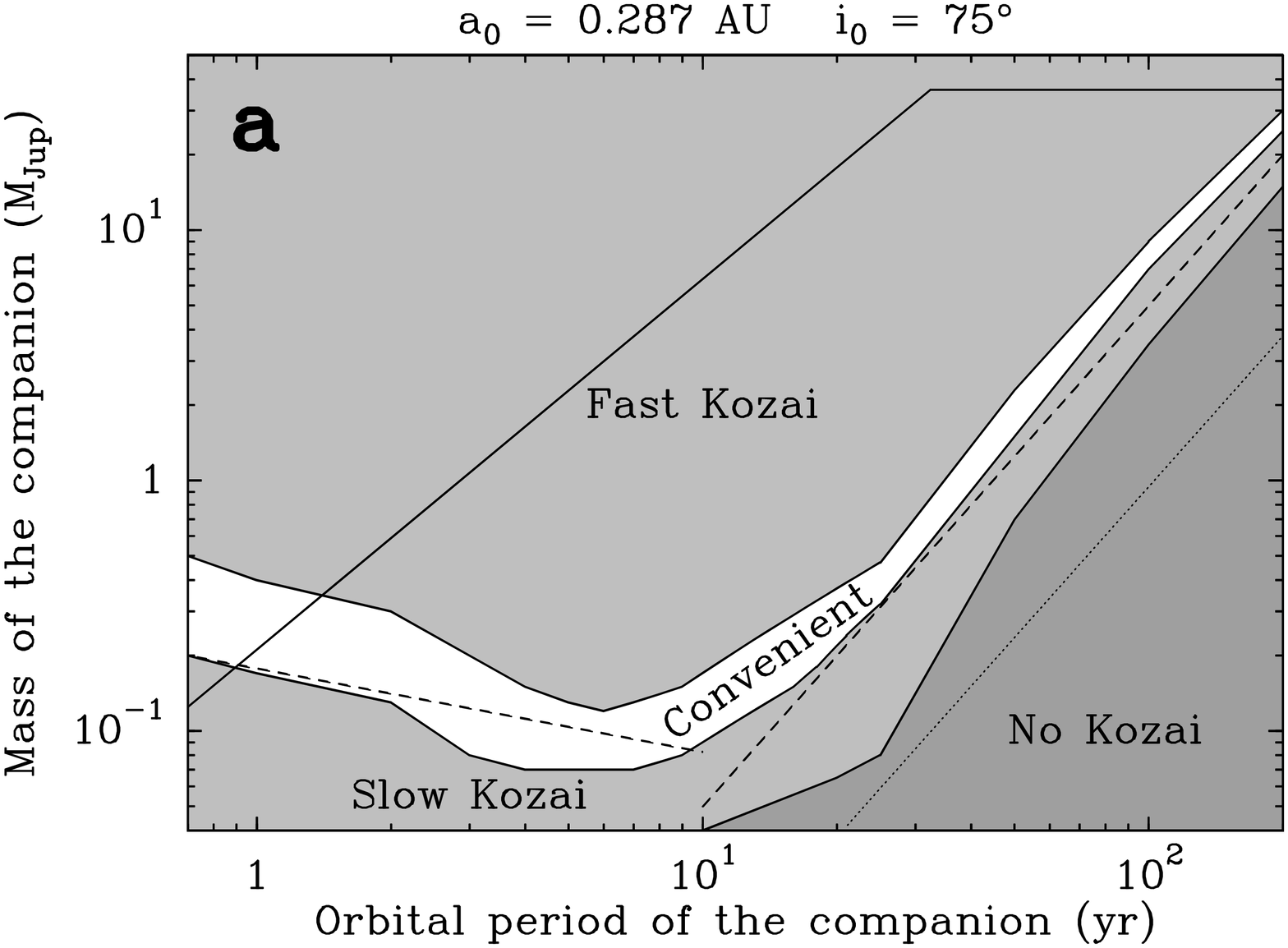} \hfil
\includegraphics[angle=-90,width=0.49\textwidth]{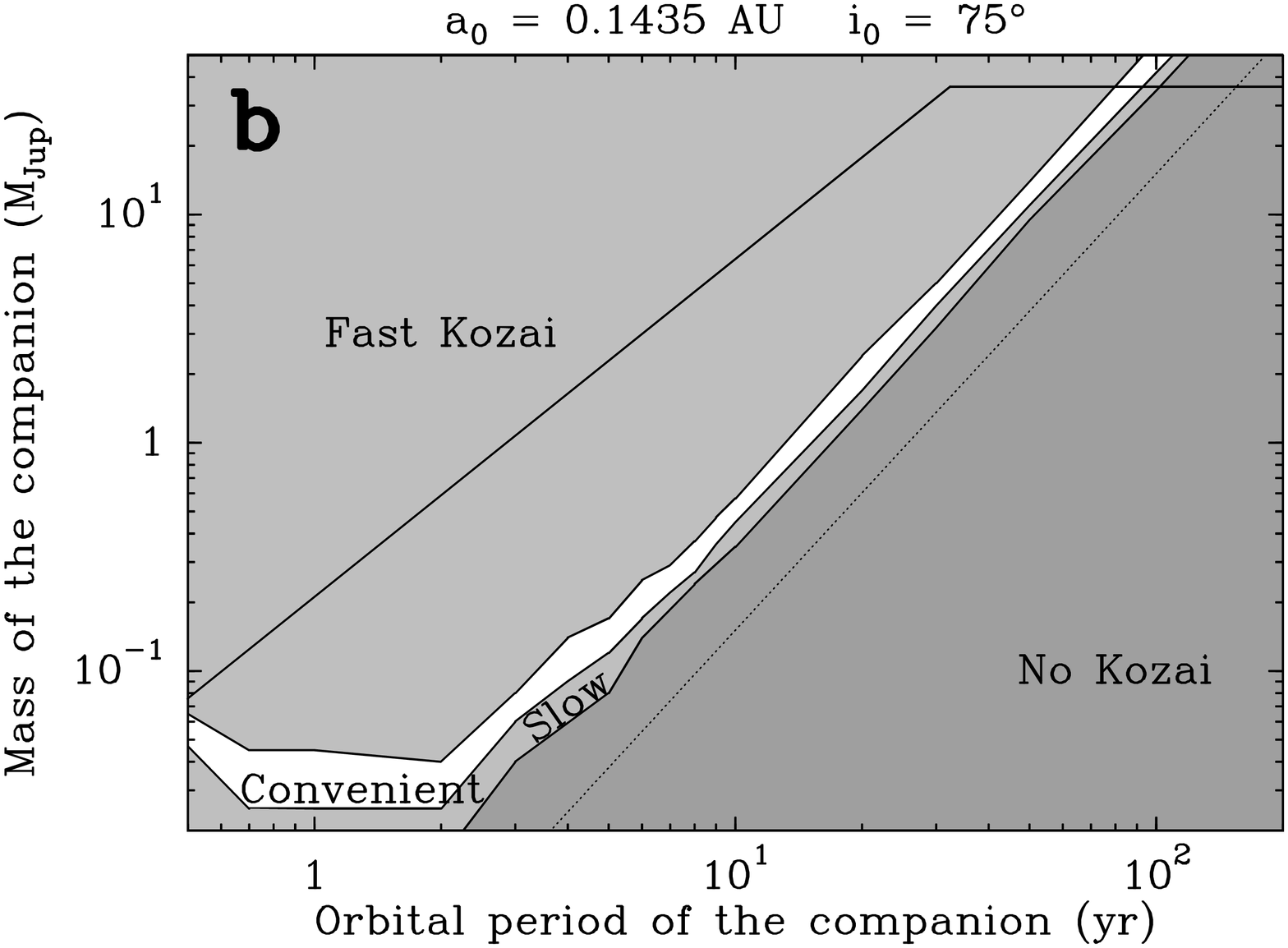}}
\vspace{0.2cm}
\makebox[\textwidth]{
\includegraphics[angle=-90,width=0.49\textwidth]{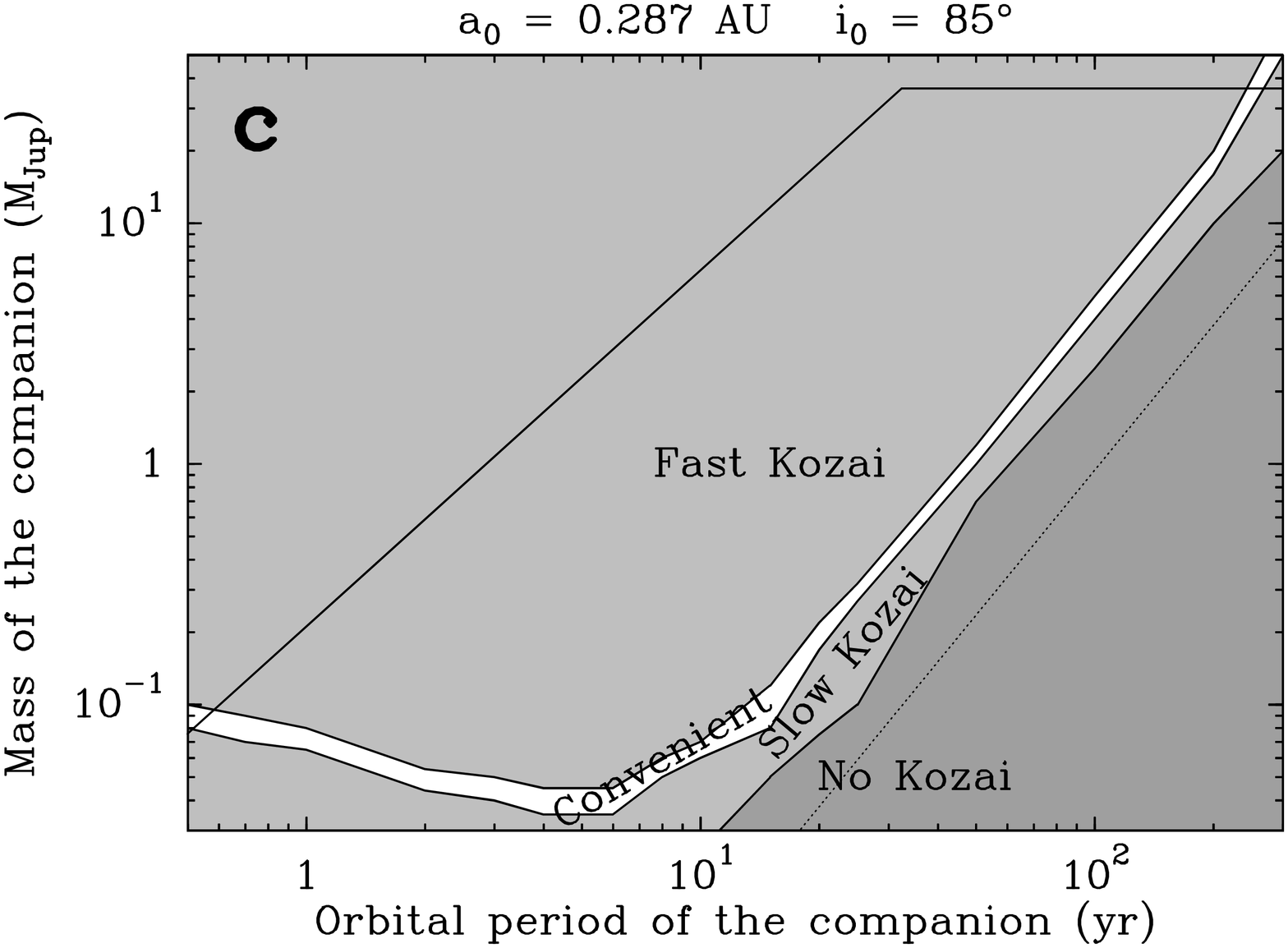} \hfil
\includegraphics[angle=-90,width=0.49\textwidth]{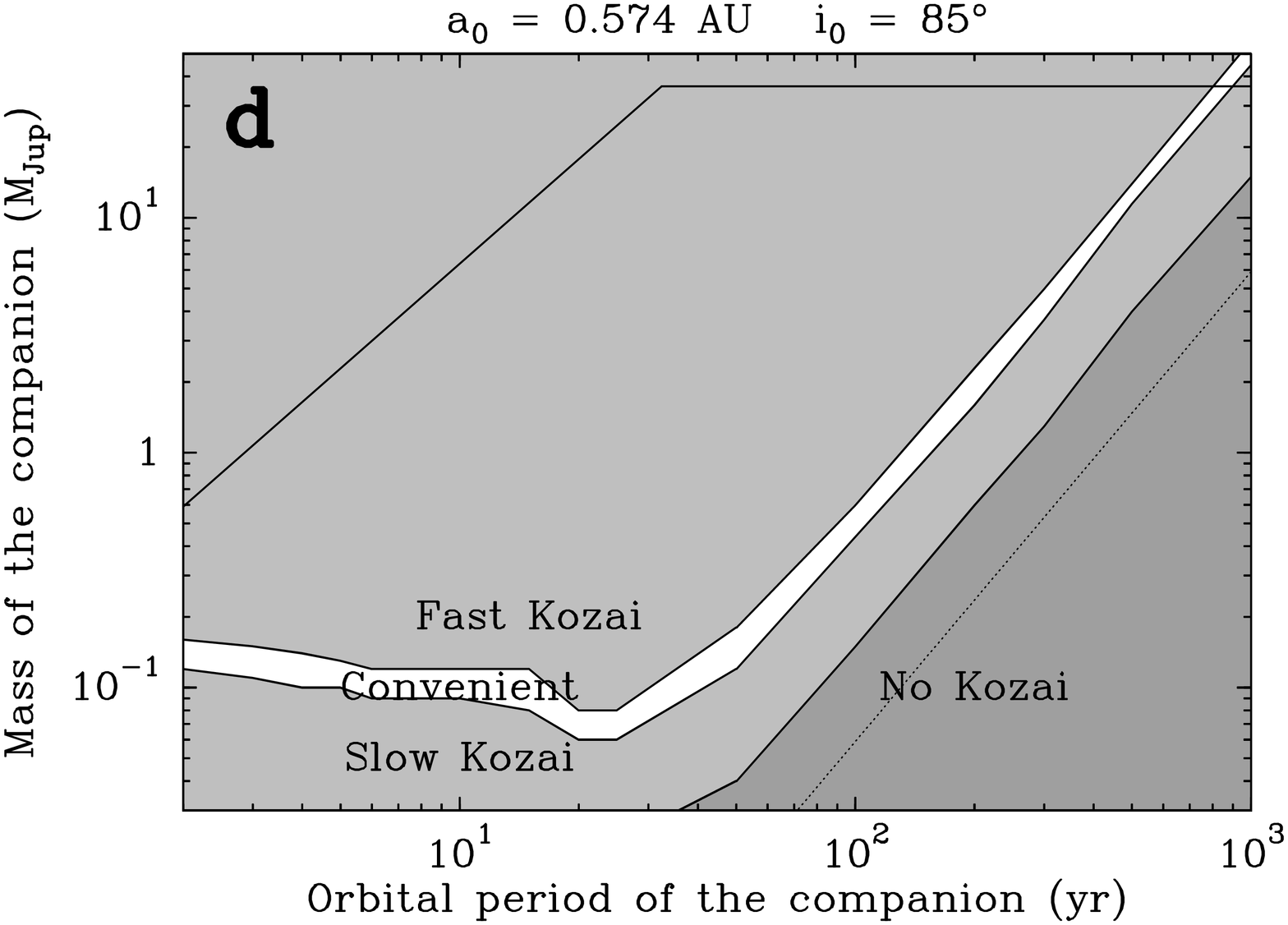}}
\caption[]{Result of the exploration of the parameter space for four
$(a_0,i_0)$ combinations. 
\textbf{a:} $a_0=0.287\,$AU, $i_0=75\degr$;
\textbf{b:} $a_0=0.1435\,$AU, $i_0=75\degr$;
\textbf{c:} $a_0=0.287\,$AU, $i_0=85\degr$;
\textbf{d:} $a_0=0.574\,$AU, $i_0=85\degr$.
In each plot, the upper observational
limits (radial velocities and imaging) are represented as a solid
line; the oblique dotted line corresponds to the limit
$t_\mathrm{GR}=t_\mathrm{Kozai}$ as given by \citet{mat09}.In the
upper left plot, the dashed lined are $m'\propto T'{-1/3}$ and
$m'\propto T'{^2}$ power laws (see text).}
\label{simdata}
\end{figure*}
Figure~\ref{simdata} shows a summary of the result of this exploration of the
parameter space. Each plot corresponds to the use of a fixed set of
values ($a_0,i_0$) where we let the other parameters $(T',m')$
vary. According to the simulation results, we divide the $(T',m')$
plane into different regions depending on the resulting value of
$\ttr$. Regions marked ``Fast Kozai'' correspond to cases where
$\ttr<1\,$Gyr. In these regions, the Kozai migration process should
have ended a long time ago, and we would expect \glb\ to be
synchronized and circularized today. This situation logically applies
to massive and/or close-in perturbers (top-left parts of the panels),
when the disturbing action is stronger and the period of Kozai cycles
smaller.  Conversely, towards the bottom parts of the panels
(small and/or distant perturbers), we have regions marked as ``Slow
Kozai'', thus corresponding to $\ttr>10\,$Gyr. In such cases, $\ttr$
is just too long to allow \glb\ to lie in the second phase of the
Kozai migration process today. This cannot match the observation, as
during the first phase of the evolution (Kozai cycles), the semi-major
axis only undergoes a very small decrease an cannot reach the present
day value. We also note in some cases regions marked ``No Kozai''
corresponding to very low mass and distant companions. In such cases,
the perturbing action of the companion is so weak that GR (even at the
considered distance) is strong enough to prevent the onset of Kozai
cycles. Thus \glb's eccentricity never grows and it can never reach
its present-day location. To illustrate this fact, we add to all plots in 
Fig.~\ref{simdata} an oblique line which corresponds in each case 
to $t_\mathrm{GR}=t_\mathrm{Kozai}$, where $t_\mathrm{GR}$ and $t_\mathrm{Kozai}$
are characteristic times for GR and for Kozai oscillations respectively.
These timescales read \citep{mat09}
\begin{eqnarray}
t_\mathrm{GR}& = &\frac{Tc^2a\left(1-e^2\right)}{3G\left(M+m\right)}\qquad,\\
t_\mathrm{Kozai}& = & \frac{2T'^2}{3\pi T}\frac{M+m+m'}{m'}\left(1-e'^2\right)^2
\qquad.
\end{eqnarray}
Here we have computed these timescales assuming $e=0$. Note that the
real period of the Kozai cycles equals $t_\mathrm{Kozai}$ within a
factor of order unity \citep{ford00,ggdyn}. It is straightforward to
see that $t_\mathrm{GR}=t_\mathrm{Kozai}$ implies $m'\propto T'^2$
(with $m'\ll M$), hence the oblique dotted lines in Fig.~\ref{simdata}.
No Kozai oscillations are to be expected below this limit 
($t_\mathrm{GR}<t_\mathrm{Kozai}$), which is 
confirmed by our numerical exploration. The ``No Kozai'' regions extend 
indeed somewhat above the dotted lines the our plots. This is due to the 
higher order effects that are not taken into account in the above 
simple formulas.

In the middle of the fast and slow Kozai regions, a convenient strip
appears where the value of $\ttr$ allows \glb\ to lie presently in the
middle of the second phase of the Kozai migration process. Note that
for a given set of parameters, the probability of witnessing this
phase today is not small, as the second phase lasts typically
$\sim0.5\ttr$.

To all plots in Fig.~\ref{simdata} we superimpose a solid line that
corresponds to the observational upper limits of \citet{mon09} for any
additional planet to \gl. The left oblique part corresponds to the
radial velocity limit and the horizontal upper part to the imaging 
(adaptive optics)
limit. Consequently, any set of $(T',m')$ values located above this
line must be rejected, and for an initial set of parameters
($a_0,i_0$) to be possible, the convenient strip must cross the region
below the observational limits. In all plots, the lower $T'$ value (in
$x$-axis) was set according to the stability limit of the
system. Systems with smaller $T'$ values are unstable. The stability
was tested for these orbits with the unaveraged symplectic
integrator. The upper limit in $T'$ was set at the point where the
convenient strip definitely gets above the observational imaging
limit, so that any convenient companion located further away would
have been already detected. So, for any given set of initial conditions
($a_0,i_0$), the convenient values for $T'$ and $m'$ must appear on the plot.

For all displayed cases, the convenient strip appears
two-fold. Towards lower $T'$ values (close-in perturbers), the
convenient $m'$ values are a slowly decreasing function of $T'$, while
for more distant perturbers, they increase rapidly. This can be
understood with a careful analysis of the Kozai mechanism.  The
efficiency of the Kozai migration scenario depends actually on how
close \glb\ can get to the star during the first phase (hence the
maximum eccentricity it can reach) and the time it spends at high
eccentricity to allow tides to be at work. For close-in perturbers,
the process is limited by the maximum eccentricity it can reach. In
fact in Fig.~\ref{simdata}, close-in perturbers in the convenient
strip have masses comparable to that of \glb\ ($\simeq 0.074\mjup$), and have
semi-major axes only a few times more than \glb\ initially. Hence the 
initial angular momentum ratio $l'_0/l_0$ between the perturber and \glb\ is 
of order unity or a bit more. Now, due to total angular momentum conservation 
(we neglect exchanges with spins), we necessarily have
\begin{equation}
l_0^2+{l'_0}^2+2l_0l'_0\cos i_0=l^2+l'^2+2ll'\cos i\qquad,
\end{equation}
where $l'$ and $l$ stand for the angular momenta of both bodies at
peak eccentricity, and $i$ for the mutual inclination in the same
phase. Assume now $l'=l'_0$ (the perturber is not
affected) and $l=l_0\sqrt{1-e^2}$ (the
semi-major axis is only little affected in the first phase). 
This equation simplifies to 
\begin{equation}
2\frac{l'_0}{l_0}\left(\cos i_0-\sqrt{1-e^2}\cos i_0\right)+e^2=0
\end{equation}
Solving this equation for $e$ shows that for moderate $l'_0/l_0$ values, the
eccentricity cannot go beyond a maximum value that is significantly below
1, while for larger value, any value as close possible to 1 is
allowed. In other words, the angular momentum of \glb\ cannot reach
zero because that of the perturber cannot grow to account for the
whole angular momentum of the system. Let us assume for instance
$i_0=75\degr$, $i=30\degr$ (typical values) and $l'_0/l_0=2$.  We
derive $e\simeq0.74$, which is high, but significantly below 1 to
avoid \glb\ to be efficiently tidally braked in peak eccentricity
phases.  The efficiency of the Kozai migration process is therefore
limited by the maximum eccentricity value than can be reached, hence
by the value of $l'_0/l_0$. Similar $\ttr$ correspond to similar
$l'_0/l_0$ values. Assuming now $l'_0\propto m'T'^{1/3}$ and
$l_0\propto mT^{1/3}$ (low initial eccentricities), a constant $\ttr$
leads $m'\propto T'^{-1/3}$, hence the decrease of the convenient strip
towards close-in perturbers.

For more distant perturbers, the $l'_0/l_0$ ratio is large enough not
to limit the peak eccentricity of the Kozai cycles. The efficiency
of the process is now limited by the amount of time spent in high
eccentricity phases, as tides are only active there. We thus expect
$\ttr\propto t_\mathrm{Kozai}$, as the longer $\ttr$, the less time is spent
within a given time span in high eccentricity phases. As 
$t_\mathrm{Kozai}\propto (M/m')\times(T'^2/T)$, a fixed
value for $\ttr$ will correspond to a constant ratio $T'^2/m'$. This
explains the increase of the convenient strip towards distant
perturbers. A more distant perturber just needs to be more massive to
generate the same Kozai migration as a closer one.

To illustrate this simple analysis, we have drawn in
Fig.~\ref{simdata}a two dashed lines corresponding to arbitrary
$m'\propto T'^{-1/3}$ and $m'\propto T'^2$ power laws. We note that
these power laws described above as well followed in the simulations
for both regimes, as the convenient strip quite faithfully remains
parallel to the dashed lines sketched. To summarize, the convenient
 strip roughly follows a dual power law regime, with
$m'\propto T'^{-1/3}$ for small $T'$ and $m'\propto T'^2$ for large
$T'$. Its exact location in  ($T',m'$) space depends on the 
initial set of parameters ($a_0,i_0$), as can be seen from 
Fig.~\ref{simdata}.

We also note that in all cases, the convenient strip appears quite
narrow, although this effect is enhanced on Fig.~\ref{simdata}
by the logarithmic scale. In fact, for a given period $T'$, Kozai migration
is very sensitive to the mass $m'$ of the perturber, so that $m'$ must
assume a well defined value within a factor $\sim 1.5$ to allow $\ttr$
to fall in the right range. However, for a wide range of initial values sets 
$(a_0,i_0)$ (see discussion below), there is a
convenient strip in ($T',m'$) which is compatible with the 
observational constraints, so that the probability of the scenario is not 
small. 
\subsection{Discussion}
We come now to compare the location of the convenient strip between
various configurations. If we compare Figs.~\ref{simdata}a and
~\ref{simdata}c (same $a_0$ but different $i_0$), we note that for
$i_0=85\degr$ the convenient strip is located lower than for
$i_0=75\degr$. Here again this may be easily understood. With a larger
$i_0$ the Kozai mechanism is stronger, so that \glb\ more easily gets
a high peak eccentricity and the tides act more efficiently. Hence a
less massive perturber is required to achieve the same result.

If we compare configurations with same $i_0$ but different $a_0$
(Figs.~\ref{simdata}a and b on the one hand, and c and d on the other
hand) we can see how the convenient strip moves in the ($T',m'$) diagram 
if we vary $a_0$. We see that increasing $a_0$ brings the convenient strip
up in the decreasing part of the diagram ($m'\propto
T'^{-1/3}$), and down in the increasing part of the diagram ($m'\propto T'^2$).
If for instance we superimpose Figs.~\ref{simdata}a and b, the convenient strip
for $a_0=0.1435\,$AU falls below that for $a_0=0.287\,$AU for small $T'$,
then crosses it and get above for large $T'$.
This can be explained. In the decreasing part, the efficiency of the Kozai
mechanism is controlled by the angular momentum ratio $l'_0/l_0$. for
a larger $a_0$, $l_0$ is larger and the angular momentum ratio is
smaller, so the Kozai mechanism is less efficient. We thus need a
larger $l'_0$ (and hence a larger $m'$) to get the same results. 
This
is why the convenient strip is brought up for a larger $a_0$. 
In the increasing part of the diagram, the
Kozai mechanism is controlled by $t_\mathrm{Kozai}$. As we know that
$t_\mathrm{Kozai}\propto T^{-1}$, a larger $a_0$ leads naturally to a
smaller $t_\mathrm{Kozai}$ and to a more efficient Kozai migration
process. Hence a less massive companion (larger $m'$) for the same
$T'$ is required to achieve the same $t_\mathrm{Kozai}$. This is why
the convenient strip is brought down if we increase $a_0$.

This regime tends nevertheless not to be valid for all values of
$a_0$.  For larger initial orbits, the whole convenient strip is
shifted upwards if we increase $a_0$. For instance, for
$a_0=0.574\,$AU and $i_0=75\degr$ (configuration not displayed in
Fig.~\ref{simdata}), the convenient strip is located much higher than
with $a_0=0.287\,$AU, so that is it barely compatible with the
observational constraints. Similarly, with $i_0=85\degr$, the
convenient strip is shifted upwards for $a_0\ga 0.6\,$AU. This
behaviour is due to the fact that the efficiency of the tides are
related to the minimum periastron reached rather than the maximum
eccentricity. As in the first phase, the semi-major axis of \glb\ is
only little affected, for a larger $a_0$ a higher eccentricity is
necessary to reach the required periastron range.

Now, is it possible to set absolute limits to the unknown parameters ?
Setting a lower limit to $i_0$ is easy. As described above, if we
decrease $i_0$ the convenient strip gets higher in ($T',m'$) space. At
some point it crosses the observational limits so that the considered
($a_0,i_0$) must be rejected. We tested several other runs with
$i_0=65\degr$ and $i_0=70\degr$. In all cases the convenient strip (if
any) appears above the observational limits. It is then possible to
stress that $i_0$ must be at least 70--75$\degr$. Note also that the
Kozai mechanism is symmetrical with respect to $i_0=90\degr$. We
checked that runs assuming $i_0=95\degr$ and $i_0=105\degr$ behave
exactly the same way as corresponding runs with $i_0=85\degr$ and
$i_0=75\degr$ respectively. It is thus possible to put an upper limit
to $i_0$ to 105--110$\degr$. The initial configuration of the system
must have been close to perpendicular within less than $20\degr$. 
This is of course valid for small initial eccentricity $e_0$. If $e_0$
is high, then $i_0$ may take different values. In the general case, a more 
relevant constraint must be derived for $h=\sqrt{1-e^2}\,\cos i$. 
$70\degr\la i_0\la 90\degr$ for small $e_0$ means $|h|\la 0.34$, which 
is valid irrespective of $e_0$.

It is also possible to set a lower limit to $a_0$. Looking at
Fig.~\ref{simdata}, we see that for lower $a_0$ the convenient strip
gets shifted towards left, so that at some point it should cross the
radial velocity observational limit. This occurs for $a_0\simeq
0.1\,$AU. Hence we can stress that the initial value $a_0$ must 
have been larger.

Giving an upper limit to $a_0$ is less easy. We first can notice than
choosing a larger $a_0$ implies a much more restricted compatible
range for $i_0$. As noted above, for $i_0=75\degr$, $a_0=0.287\,$AU is
possible, but $a_0=0.574\,$AU is barely excluded (too high convenient
strip). For $i_0=85\degr$ nevertheless, we tested that even fairly
large $a_0$ values (up $\sim 10\,$AU) we can find a convenient strip
in $(T',m')$ space that falls still below the observational limits. We
may thus set an absolute upper limit to $a_0\simeq 10\,$AU. However,
we must point out that for such values, we are in an extreme Kozai
regime with a very high peak eccentricity. Even if such a situation
cannot be rejected, whether the onset of such a regime with nearly
orthogonal orbits is likely is questionable. Hence even if the
absolute compatible range for $a_0$ is $0.1$--$10\,$AU, we may stress
that the range $0.1$--$0.5\,$AU is more probable.

As pointed out above, in all cases, the convenient strip in $(T',m')$
space to allow $\ttr$ to fall in the suitable time range is quite
narrow. This does not tell that the situation we are describing is
improbable, as suitable configurations are possible for a wide range
of initial configurations. This tells that for any given set of values
$(a_0,i_0,T')$ the mass of the companion $m'$ must fall around a well
defined value within a factor $\sim 1.5$. We nevertheless note that
taking larger $a_0$ values leads to generate narrower convenient
strips. This shows up already in Fig.~\ref{simdata}d
($a_0=0.574\,$AU), and it is even more striking for larger $a_0$. This
is another reason why we think that a configuration with a moderate
$a_0$ is more likely.

Our parametric study also shows that the evolution we describe is not
the only one possible. First the initial inclination $i_0$ must be
high to have a Kozai migration. Second, if the perturber's mass $m'$
is too small, there is no Kozai migration, and if it is too high,
Kozai migration is extremely rapid.  Kozai
migration was also suspected to account for the misaligned systems
that have been measured from Rossiter-McLaughlin effect. But among all
the systems for which an obliquity with respect to the stellar axis has
been measured, only $\sim 30\,\%$ show an obliquity larger than $45\degr$
\citep{heb11}. This shows that Kozai migration is not a universal process. 
The limited size of the convenient strip in our parametric study supports 
this idea. Systems like
\gl\ with eccentric close-in planets, although not unrealistic, are not 
expected to be numerous.
\section{Conclusion}
The Kozai migration scenario is a generic process that causes inward
migration of planets in non coplanar planetary systems. This evolution
is characterized by two phases, a first one with perturbed Kozai
cycles, and a second one with a more drastic decay of the semi-major
axis. In any case, the characteristic time for this evolution can be
considerably longer than the standard tidal circularization time,
providing this way a solution to the issue of the high present
eccentricity of \glb. If we combine our study and that of
\citet{mon09}, various suitable initial configurations can be found. They all
require a high initial mutual inclination ($\ga 75\degr$), an initial
orbit for \glb\ several times wider than today, perturber masses
ranging between less than $0.1\,\mjup$ and $50\,\mjup$, with orbital
periods ranging between a few years up to several hundreds. This model 
is also generic enough to be invoked
to explain some other puzzling cases of high eccentricity close-in
exoplanets. But for a
given set of initial conditions, the orbital period and the mass of
the perturber must assume a fairly well constrained relationship,
otherwise the suspected process is either too rapid or
nonexistent. This could then explain why system like \gl\ are not numerous.
 
Note that the constraint derived for $i_0$ is valid for small initial
eccentricity $e_0$. The real condition is $|h|\la 0.3$, which can be
achieved either with low $e_0$ and high $i_0$, but also with low $i_0$
and high $e_0$. For small $i_0$, $|h|\la 0.3$ implies $e_0\ga 0.95$.
Although this is actually just a matter of fixing the starting point
on the Kozai cycles, this has a different meaning for what concerns
the formation of such a planetary system. Planetary systems are 
usually thought to form
in circumstellar disk with initial coplanar configurations 
\citep[see e.g. theoretical studies by][]{pie08,cri09}. 
The only way to imagine an
initially inclined configuration is to assume that both planets formed
independently and were assembled further to build the
\gl\ system. This could happen for instance if \gl\ was initially
member of a multiple stellar system. Such systems are indeed rarely
coplanar. Alternatively, the system could have formed coplanar but 
the inner planet could have been driven to high eccentricity. There are 
many ways to achieve this. This can be done  
by planet-planet scattering events or chaotic 
diffusion. \citet{lask96} indeed showed (with application to Mercury, see also
\citet{wu11}) that in any planetary system, the inner planet may be 
subject to drastic eccentricity increases due to secular chaos. Similarly, 
recent work on the so-called Nice model for the formation of the Solar System 
\citep{tsig05,lev11} have demonstrated the role of planet-planet 
scattering among the giant planets. Mean-motion 
resonances can also drive inner bodies to high eccentricities. This mechanism 
was invoked to explain the suspected Falling Evaporated Bodies (FEB) scenario 
in the $\beta\:$Pictoris system \citep{bm96,bm00}.

In conclusion, an initially nearly coplanar but eccentric
configuration of the system appears more realistic than a highly
inclined one. Note that this does not change anything to the further
evolution described in this paper, as Kozai cycles cause the system to
rapidly oscillate between these two extreme configurations. Even if
the system is initially nearly coplanar, it evolves to highly inclined
within half a Kozai cycle.  The relevant issue is actually the
definition of the starting point.  According to \citet{bat11}, in all
systems potentially subject to Kozai resonance, Kozai cycles are first
frozen as long as self gravity and strong planet-planet interactions
induce a large enough apsidal precession. But at some point when the 
system relaxes, Kozai cycles can start, which constitutes 
the ``starting point'' in our scenario.  

The mean-motion resonance model that applies to the FEB scenario can
also provide a valuable starting point. The only requirement is that
both planets are initially in a mean-motion resonance like 4:1 ou 3:1,
with a small initial mutual inclination (a few degrees) and a small
eccentricity ($\ga 0.05$) for the outer, more massive planet. If this
is fulfilled, the inner body is subject to an eccentricity increase
virtually up to $\sim 1$. It was shown by \citet{bp3d} that whenever
it reaches the high eccentricity state (the FEB state in the case of
$\beta\:$Pictoris), it undergoes high amplitude inclination
oscillations that can be viewed as Kozai cycles inside the mean-motion
resonance. This is a kind of resonant version of our model that could
constitute a full coherent scenario driving the system from a coplanar
and circular configuration to strong Kozai resonance.

The reality of this scenario will nevertheless be difficult to test 
observationally. Even if the two planets are initially in mean-motion 
resonance, as soon the semi-major axis of \glb\ starts to evolve 
thanks to tidal 
dissipation, they leave the resonance. The present day orbital 
configuration keeps no memory of the initial resonance.
Assuming that we are today in the middle of the second phase
with a still significant eccentricity and an already reduced
semi-major axis (by a typical factor 10), we are now unable 
to distinguish between initially resonant and non-resonant configurations. 
We also note from Fig.~\ref{1a_long} that
the present day mutual inclination between the two planets should be in 
any case $\sim 20\degr$, which makes it difficult also to distinguish 
observationally between coplanar \citep{bat09} and Kozai models 
(this work). However, as explained above, the exact characterization of the 
apsidal behavior of the planets can lead to constraints on their mutual 
inclination. Basically, if $\varpi-\varpi'$ is close to 0 or $\pi$, this 
would constitute a strong clue in favor of the coplanar 
model of \citet{bat09}. Otherwize, this would indicate an inclined 
configuration.

Before that, a crucial issue is to continue the radial velocity and
photometric follow up of \gl. If we confirm the presence of a more
distant companion, and if we are able to significantly constrain its
orbit, then we will be able to test the proposed scenarii into more
details. This would not only resolve the eccentricity paradox of \glb,
but it would also provide clues towards the system's past evolutionary
history.
\begin{acknowledgements}
All computations presented in this paper were performed at the \textsl{Service 
Commun de Calcul Intensif de l'Observatoire de Grenoble} (SCCI) on the 
super-computer funded by \textsl{Agence Nationale pour la Recherche} under 
contracts ANR-07-BLAN-0221, ANR-2010-JCJC-0504-01 and ANR-2010-JCJC-0501-01.
We gratefully acknowledge financial support from the French \textsl{Programme
National de Plan\'etologie} (PNP) of CNRS/INSU. We thank our referee, 
Konstantin Batygin, for valuable suggestions to improve the discussion 
of the paper.
\end{acknowledgements}
\end{document}